\documentclass{article}

\usepackage{arxiv}

\usepackage[utf8]{inputenc} % allow utf-8 input
\usepackage[T1]{fontenc}    % use 8-bit T1 fonts
\usepackage[hidelinks]{hyperref}      % hyperlinks
\usepackage{url}            % simple URL typesetting
\usepackage{booktabs}       % professional-quality tables
\usepackage{amsfonts}       % blackboard math symbols
\usepackage{amsmath}
\usepackage{nicefrac}       % compact symbols for 1/2, etc.
\usepackage{microtype}      % microtypography
\usepackage{lipsum}		    % Can be removed after putting your text content
\usepackage{graphicx}
\usepackage{natbib}
\usepackage{doi}
\usepackage{rotating}
\usepackage{makecell}
\usepackage{xcolor}
\usepackage[normalem]{ulem}
\usepackage{soul}
\usepackage{longtable}
\usepackage{array} 
\usepackage{booktabs}
\usepackage{enumitem} %for controlling enumerate
\usepackage{comment}
\usepackage{caption}
\usepackage{adjustbox}
\usepackage{floatrow}

\graphicspath{ {./Figures/} }

\newcommand{\pkg}[1]{{\normalfont\fontseries{b}\selectfont #1}}
\let\proglang=\textsf
\let\code=\texttt

\usepackage{floatrow}
\usepackage{caption}
\usepackage{subcaption}
\usepackage[rightcaption]{sidecap}
\usepackage{longtable,booktabs,array}

\title{Self-referentiality and asymmetric knowledge flows between journals. The case of economics}

\date{} 					% Or removing it

\author{\href{https://orcid.org/0000-0003-0293-482X}{\includegraphics[scale=0.06]{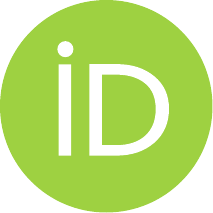}\hspace{1mm}Alberto Baccini}%\thanks{Use footnote for providing further
  \\
	Dipartimento di Economia Politica e Statistica\\
	Università degli Studi di Siena\\
	Siena, Italy \\
	\texttt{alberto.baccini@unisi.it} \\
	\And \href{https://orcid.org/0000-0002-7210-4541}{\includegraphics[scale=0.06]{orcid.pdf}\hspace{1mm}}{Carlo Debernardi}\\
	Dipartimento di Economia Politica e Statistica\\
	Università degli Studi di Siena\\
	Siena, Italy \\
	\texttt{carlo.debernardi@unisi.it} \\
	}

\renewcommand{\headeright}{}
\renewcommand{\undertitle}{}
\renewcommand{\shorttitle}{Self-referentiality and asymmetric knowledge flows between journals}

\hypersetup{
pdftitle={},
}

\makeatletter
\input{size12.clo}
\makeatother

\begin{document}

\maketitle

\begin{abstract}

This paper investigates the evolution of self-referentiality and knowledge flows in economics journals before and after the 2008 financial crisis. Using a multi-level approach, we analyze patterns at the discipline, cluster, and journal levels, combining citational measures with a classification of journals based on intellectual similarity and social proximity.
At the aggregate level, results suggest a general decline in self-referentiality, indicating increased openness across the discipline. However, this trend conceals substantial heterogeneity. At finer levels of analysis, two clusters – CORE and Finance – emerge as persistent outliers, exhibiting very high levels of self-referentiality. While Finance experienced a gradual reduction over time, the CORE shows increasing closure.
By examining reference asymmetries, we uncover a hierarchical structure of knowledge flows. The CORE operates as a central hub and net exporter of knowledge to all other clusters, particularly to the traditional core fields of economics, whereas Finance acts as a net exporter only within its own domain and remains dependent on the CORE. These asymmetries are reinforced at the level of individual journals, where a small set of top journals occupies the apex of a hierarchically ordered system of knowledge transmission.
We argue that these patterns reflect the interplay between intellectual dynamics and organizational structures, particularly the role of editorial networks in shaping access to publication and visibility. The findings suggest that, following the financial crisis, economics has experienced a process of increasing epistemic and organizational closure at its core, alongside greater openness in peripheral areas. This dual dynamic raises questions about the representativeness of top journals and the evolving structure of the discipline.

\end{abstract}


\newpage

\section{Introduction}

%\cite{Fourcade} % The Superiority of Economists
%\cite{jacobs2014} % In Defense of Disciplines
%\cite{stigler1995} % The Journals of Economics
%\cite{angrist2017, angrist}
%\cite{Cedrini}
%\cite{aistleitner2019}

%%%%%%%%%%%%%%%%%%%%%%%%%%%%%%%%%%%%%%%%%%%%%%%%%%%%%%%%%%%%%%

The degree of interdisciplinarity in economics -- or, conversely, its tendency toward intellectual self-referentiality -- has long been a subject of debate. Early quantitative studies of citation patterns suggested that economics maintains relatively limited intellectual exchange with neighboring disciplines. \citet{pieters202}, for instance, find that economists rarely build directly on insights developed in other social sciences. \citet{moody2006} show that economics journals form a distinct and tightly connected cluster within the broader network of social science publications, with strong internal citation patterns. These empirical observations have been paralleled by broader interpretative claims about the discipline’s position within the academic landscape. In particular, the perceived superiority of economics has been linked to the phenomenon of economic imperialism \citep{maki2009}, whereby economic theories and methods expand into domains traditionally studied by other social sciences while drawing comparatively little on their intellectual contributions.

\citet{Fourcade} intervened in this debate arguing that economics is uniquely insular among the social sciences. Combining several indicators -- including citation flows, publication practices, and professional hierarchies -- they show that economics exhibits unusually high levels of internal referencing and relatively limited engagement with external disciplines.

\citet{angrist} have challenged this interpretation. While acknowledging that economists frequently cite within their own discipline, they argue that economics is relatively insular, but not uniquely so when compared with other social sciences. Instead, they emphasize the substantial influence that economic research exerts on neighboring fields, highlighting the diffusion of economic methods and empirical strategies across a wide range of disciplines. From this perspective, the apparent self-referentiality of economics coexists with a growing degree of intellectual interaction, particularly through the outward transmission of analytical tools and empirical research designs. However, this perspective focuses primarily on the external impact of economics rather than on the extent to which economic research incorporates insights from other disciplines, which leaves open the question of whether intellectual exchange is genuinely reciprocal.

Recent work by \citet{Truc_2023} revisits this issue and offers evidence that partly reconciles these perspectives while also challenging some of Angrist et al.’s conclusions. Their analysis confirms that interactions between economics and other disciplines have increased over time, particularly since the 1990s. At the same time, however, they show that economics remains comparatively more inward-looking than other social sciences and that intellectual exchanges with neighboring fields are often asymmetric. In particular, economics tends to cite other disciplines less frequently than it is cited by them. Moreover, \citet{Truc_2023} find that the most prestigious journals in economics remain especially self-referential, suggesting that the discipline’s hierarchical structure plays an important role in shaping patterns of intellectual exchange.

This study contributes to this line of research by providing a more comprehensive and fine-grained empirical assessment of self-referentiality in economics. First, we rely on a large dataset covering 711 journals indexed in EconLit, which allows us to capture a substantially broader portion of the discipline than most previous analyses. Second, we move beyond treating economics as a homogeneous corpus of publications by examining citational patterns across different subfields of the discipline.

To do so, we adopt a research strategy that differs from previous studies in two respects. The first consists in considering journals as the main unit of analysis and examining whether the aggregate patterns identified in earlier research persist when the discipline is analyzed at a more granular level. The second consists in adopting a fine-grained classification of economics journals that we developed in a previous contribution \citep{baccini2025}. In this framework, journals are grouped into clusters based on the textual content of their abstracts, their shared knowledge bases, and the overlapping communities of editors and authors associated with them. This classification allows us to investigate how patterns of self-referentiality vary across different areas of the discipline.

Building on this framework, the study addresses three main research questions. First, we examine the degree of self-referentiality exhibited by different clusters of economics journals and assess how it varies when moving from the analysis of the entire corpus to clusters and then to individual journals. Second, we analyze whether citational relationships between clusters are symmetric or asymmetric -- in other words, whether some areas of the discipline primarily export intellectual influence while importing relatively little from others. Third, we investigate how these patterns have evolved over time by introducing a temporal dimension that distinguishes between the period preceding the 2008 financial crisis and two subsequent periods.

Our results confirm the high levels of self-referentiality previously documented in the literature, particularly within the core journals of the discipline, which are in line with the 80\% figure reported by \citet{jacobs2014}. At the same time, we find that interactions with other disciplines are concentrated primarily in more applied areas of economics. These patterns suggest that the apparent increase in interdisciplinarity observed in recent decades may largely reflect the expansion of applied subfields rather than a generalized transformation of the discipline as a whole. Moreover, the hierarchical organization of economics is reproduced within the structure of its citation network: journals located at the theoretical core occupy highly centralized positions, exporting references toward other areas of the discipline while importing relatively little intellectual influence in return. Finally, the engagement of applied areas with other disciplines appears to combine elements of genuine interdisciplinarity with forms of intellectual exchange that may reflect the outward diffusion of economic methods accompanied by relatively limited incorporation of external theoretical frameworks.

%%%%%%%%%%%%%%%%%%%%%%%%%%%%%%%%%%%%%%%%%%%%%%%%%%%%%%%%%%%%%%

\section{Field delineation and data}

Our delineation problem is twofold since we need a operational definition of economics, and a suitable classification of it in different sub-fields. 

In order to delineate economics as a scientific field, our starting point is a set of journals that ``experts'' classified as economics \citep{Zitt}. This set contains about 1,700 journals indexed in EconLit in 2019, and collected in the GOELD (Gatekeepers of Economics Longitudinal Database) \citep{baccini_re}. EconLit provides the best bibliographic coverage of the economics literature \citep{Gusenbauer}, and selective representation of adjacent fields, including business and statistics. This list reflects economics community's own conception of its publishing domain \citep{Cherrier}. Delineated in this way, the economics field is larger than as defined in similar analyses, which are often limited to very small subsets of ``top'' journals \citep[e.g.][]{angrist, galiani}. An exception to the small scale analysis is \citet{Truc_2023} where economics is represented as a list of 384 journals classified as economics by the US National Science Foundation.

As for the classification in sub-fields of the domain of economics, we adopt the clustering of EconLit journals proposed in a previous analysis \citep{baccini2025} which resulted in an extensive map of the economic discipline, partitioned according to both social and intellectual criteria. This map covers a total of 1,075 journals for which bibliographic metadata of articles are available on OpenAlex. 

More specifically, \citet{baccini2025} classified journals into communities by combining information about their topical content (i.e. the similarity of the abstracts of the papers they published), shared knowledge-base (i.e. bibliographic coupling), shared authorships and interlocking editorship. These four dimensions were integrated using a technique called Similarity Network Fusion to produce a unified similarity network from which journal communities were identified using a consensus clustering approach.  More details about the data and methods are available in the original study. 

In particular, this paper considers EconLit journals that in the previous analysis were classified in clusters composed of at least ten journals. These clusters were 34 out of 74. The final sample consists of 707 citing journals. 

The clusters of citing journals, as reported in Table \ref{tab:communities}, appear to be heterogeneously formed. Foremost among them is the CORE cluster (Cluster 3), which assembles 33 journals. The CORE functions as a prestige-filtered distillation of the discipline as a whole: it includes the so-called Top-Five, all journals published by the American Economic Association, and leading journals in macroeconomics, labor economics, and development, together with journals issued by the most relevant international economic institutions (the IMF and World Bank). Other clusters are grouped around shared theoretical orientations, exemplified by the Post-Keynesian and Heterodox Approaches (cluster 4) -- which gathers general interest journals sharing a non-neoclassical approach -- or by the History of Economic Thought and Methodology (cluster 25). The Economics of Innovation and Technology (cluster 15) blends subject-based and method-based specialization. Microeconomic Theory (cluster 19) stands as a clear instance of a theory-based sub-field. Most remaining groups represent applied subject-matter specialties, bringing together journals focused on specific areas such as agricultural, environmental, industrial, labour, real-estate, transport, and health economics. Other groups gather journals from contiguous disciplines indexed in EconLit, such as Accounting (cluster 8), Operations Research (cluster 17), or Statistics -- this latter aggregated with econometrics (cluster 6).

\begin{table}[]
\caption{Clusters of citing journals and their size. The clustering is developed in \citet{baccini2025}. Only clusters with 10 journals or more are considered.}
\small
\begin{tabular}{clc}
\toprule
\textbf{Cluster}& \textbf{Cluster label} & \textbf{N. of Journals} \\ 
\midrule
1 & European Applied Social Sciences: Regional Economics, Policy, and Management & 37 \\ 
  2 & Finance & 36 \\ 
  3 & CORE & 33 \\ 
  4 & Post-Keynesian and Heterodox approaches & 33 \\ 
  5 & Global Economics  and  Islamic Finance: Policy, Trade, and Regional Dynamics & 31 \\ 
  6 & Econometrics and Statistics & 30 \\ 
  7 & Agricultural economics & 30 \\ 
  8 & Accounting & 28 \\ 
  9 & Regional and Spatial economics & 28 \\ 
  10 & Emerging and transitional economies & 26 \\ 
  11 & International Cooperation and Development & 25 \\ 
  12 & Finance [topics] & 25 \\ 
  13 & Business studies & 25 \\ 
  14 & Economics of Energy, Resources and Environment & 24 \\ 
  15 & Economics of Innovation and Technology & 22 \\ 
  16 & Economics and finance & 22 \\ 
  17 & Operations research & 18 \\ 
  18 & Development economics & 18 \\ 
  19 & Microeconomic theory & 17 \\ 
  20 & Labour Economics & 16 \\ 
  21 & Industrial Organization & 15 \\ 
  22 & Topics in economics & 14 \\ 
  23 & Health Economics & 15 \\ 
  24 & Industrial relations & 15 \\ 
  25 & HET and Methodology & 14 \\ 
  26 & Spanish-language & 13 \\ 
  27 & Housing and Urban studies & 14 \\ 
  28 & Asia-Eurasia: Energy, Trade, and Emerging Market Development & 13 \\ 
  29 & Economic history & 12 \\ 
  30 & Public economics, Public finance & 13 \\ 
  31 & Transport economics & 12 \\ 
  32 & Real estate and Housing economics & 12 \\ 
  33 & Development [Asia] & 11 \\ 
  34 & Turkish journals & 10 \\ 
  \bottomrule
\end{tabular}
\label{tab:communities}
\end{table}

Since, this study builds upon the dataset and findings of the prior analysis \citep{baccini2025}, it adopts the same three time windows: 2006–2008, 2012–2014, and 2019–2021. These periods represent, respectively, the interval before the 2008 financial crisis, the immediate post-crisis years, and a later phase more distant from the event.  This periodization permits to study the effects, if any, of an external shock over the self-referentiality and knowledge flows of the discipline. 

The data collection started from the list of journals and their clusters, and expanded the original data collection by retrieving metadata concerning the 1,939,807 unique references of their 258,297 published articles. This further step of data collection from the OpenAlex API \citep{priem2022} was performed in November 2025 and, given the volume of data involved, was possible thanks to the kind support given us by the OpenAlex team, which provided an API key with higher rate limits.

The size of the dataset is described in Table \ref{tab:dataset_description}. As anticipated, the ``Citing journals''  are the ones belonging to the 34 clusters we are considering. The articles published in these journals are the ``Citing articles'' for which we have collected the cited references. The ``Total citations'' indicates the total number of references contained in the citing articles. These total citations are directed to ``Cited items'', i.e. to articles, books, conferences or repositories cited at least once in the bibliographical references of the citing articles. Finally, the sets of ``Cited journals'' are composed by the journals that appear at least once in the metadata of the cited items. 

As for the cited journals, they are considered inside the domain of economics if they are indexed in EconLit database, otherwise they are considered non-economics journals. 

%When it is necessary to distinguish this dataset from other datasets, we will refer to it as "EconLit" for clarity. 

\begin{table}[ht]
\centering
\begin{tabular}{cccccc}
  \toprule
Period & Citing & Citing & Total & Cited &Cited   \\
 & journals & articles & citations & items & journals  \\
  \midrule
2006-2008 & 571 & 63,174 &1,170,919  & 470,860 & 11,750 \\ 
2012-2014 & 686 & 86,116 & 2,141,243  & 780,170  & 17,576 \\ 
  2019-2021 & 665 & 109,007 & 3,821,557  & 1,313,210 & 27,088 \\ 
   \bottomrule
\end{tabular}
\caption{Size of the dataset of EconLit journals. Citing journals: number of economics journals belonging to the 34 clusters. Citing articles: number of articles from these journals for which references were collected. Total citations: total number of references in citing articles. Cited items: number of articles, books, conferences, repositories cited at least once. Cited journals: number of journals appearing among cited items.} 
\label{tab:dataset_description}
\end{table}

\section{Indicators of self-referentiality and reference asymmetry}

Our main object of observation is the knowledge base of economics journals, as represented by their cited references. This perspective is complemented by the other side of the coin: the citations received by economics journals. Our database does not allow us to obtain a complete picture of the citations received by economics journals, as it omits citations coming from journals that are not included in our list of citing journals. Nevertheless, it enables an overall assessment — within economics — of the impact of the knowledge produced by various clusters and journals.

We proxy self-referentiality using four different types of indicators, calculated at three levels of granularity: field, cluster, and journal. These four types correspond to different definitions of self-reference adopted in the analysis. All indicators are computed as the proportion of each specific type of self-references to the total number of references calculated at the given level of granularity. More precisely, each indicator is computed as:

\begin{equation}
I_{t,g} = \frac{S_{t,g}}{R_g}
\end{equation}

where \( t \in \{1,2,3,4\} \) indexes the four types of self-references, \( S_{t,g} \) is the number of self-references of type \( t \) at granularity level \( g \), and \( R_g \) is the total number of references at the same level \( g \), with \( g \in \{\text{field}, \text{cluster}, \text{journal}\} \).

The first type of indicator measures \textit{journal self-referentiality} and is defined on the basis of \textit{journal self-references}; that is, references from articles in a journal to other articles in the same journal. Journal self-references indicates the closure of a journal on the knowledge it has produced. 

The second type of indicator measures \textit{within-cluster referentiality} considering within cluster references; that is, references from journals of a cluster to articles published in journals within the same cluster. Within-cluster referentiality indicates the closure of journals or clusters on the knowledge produced within their own cluster. 

We label the third type of indicator as \textit{in-any-cluster referentiality}; it leverages references to articles published in journals of any of the 34 clusters. Conceptually, these indicators operate as if the set of clusters of the citing journals represents the economics domain, thereby offering a first -- arguably underestimated --  proxy of within-field or disciplinary closure. 

Finally, the fourth type of indicator measures \textit{within-economics referentiality} and is based on references to articles published on EconLit indexed journals. This indicator computed at field level represents the complement to 1 of the ``Citations Outside Category'' metric used by \citet{Truc_2023} to assess interdisciplinarity and is thus directly comparable to it. 

When we go down to the cluster and journal level, it is possible to integrate the self-referentiality investigation with indicators that measure the impact of each cluster or journal on other clusters or journals within economics. Here, the focus shifts from references to citations received by clusters or journals from other clusters and journals in economics. To this end, we compute an indicator for two different levels of granularity, cluster and journal, as the proportion of self-citations of a cluster or journal over the total citations received from the set of citing journals. We refer to this indicator as \textit{self-impact}. More precisely, this indicator is computed as:

\begin{equation}
SI_{g} = \frac{SC_{g}}{C_g}
\end{equation}

where \( SI_{g} \) is the self-impact at granularity level \( g \), \(SC_g \) is the number of self-citations of \( g \), that is citations generated by \( g \) itself, and \(C_g \) is the total number of references received by \( g \), with \( g \in \{\text{cluster}, \text{journal}\} \). Naturally, $1-SI_g$ measures the impact of that cluster or journal on other citing clusters or journals.

In addition, we are interested in the knowledge flows among clusters and journals. In order to explore these flows we use data about references to compute indicators of reference asymmetry among clusters and journals. To this end, building on previous scientometric contributions developed from indicators of revealed comparative advantage \citep[for all][]{schubert}, we propose a reference asymmetry index (RA). Given two clusters or two journals $i$ and $j$, it is defined as:  
\begin{equation}
RA_{i,j} = \frac{R_{i,j}}{R_i} - \frac{R_{j,i}}{R_j}, 
\end{equation}
where $R_{i,j}$ indicates the number of references from $i$ to knowledge published by $j$; $R_{j,i}$ indicates the number of references from $j$ to $i$; and ${R_i}$ and ${R_j}$ the total number of references made by each cluster or journal. By normalizing each flow by the total citations of the citing cluster or citing journal, the index accounts for differences in size, measuring the \textit{intensity} of the relationship rather than absolute citation volumes.

The interpretation of the index is straightforward: a positive value of \( RA_{i,j} \) indicates that the scientific production of cluster or journal \( j \) holds greater relative importance for cluster or journal \( i \) than vice versa, suggesting an asymmetric attraction where \( i \) depends more on \( j \) as a source of knowledge. Conversely, a negative value signals the opposite pattern, with \( j \) showing a stronger relative orientation toward \( i \). When \( RA_{i,j} \) is close to zero, the scientific attraction between the two clusters or journala is essentially symmetric and mutually balanced. Notably, the RAI is antisymmetric by construction, satisfying \( RA_{i,j} = -RA_{j,i} \). 

\section{Data analysis}

As anticipated, our main object of observation are the cited references interpreted as the knowledge base of economics journals. The first basic information concerns the composition of this knowledge base in terms of the types of editorial outlets in which it appears. Cited references are classified according to OpenAlex typologies of outlets in four categories: Journals, Books, Conferences and Repositories.

Table \ref{tab:outlet_types} reports the absolute values and the shares of references directed toward these four different types of publications. 
The knowledge used by economics journals is mainly published in academic journals. The share of references to articles is growing and at the end of the period accounted for 95\% of all references. Books reduced their presence from 5\% to 3.5\%. Conferences, unlike other sectors, are of negligible importance, as are repositories and other residual channels. 

Given that journal articles are easily classified by attributing them to the publishing journal and that they represent the overwhelming majority of our dataset, we focus our attention on articles and exclude other materials from the following analysis.

\begin{table}[ht]
\centering
\begin{tabular}{lcccccccc}
  \toprule
  & \multicolumn{2}{c}{Journals} & \multicolumn{2}{c}{Books} & \multicolumn{2}{c}{Conferences} & \multicolumn{2}{c}{Repositories} \\
  \cmidrule(lr){2-3} \cmidrule(lr){4-5} \cmidrule(lr){6-7} \cmidrule(lr){8-9}
  Period & Cit. & Share & Cit. & Share & Cit. & Share & Cit. & Share \\
  \midrule
  2006-2008 & 1,082,778 & 0.925 & 57,632 & 0.049 & 1,775 & 0.002 & 28,734 & 0.025 \\
  2012-2014 & 2,008,033 & 0.938 & 86,190 & 0.040 & 2,337 & 0.001 & 44,683 & 0.021 \\
  2019-2021 & 3,634,926 & 0.951 & 132,669 & 0.035 & 3,592 & 0.001 & 50,370 & 0.013 \\
  \bottomrule
\end{tabular}
\caption{Types of outlets in which the knowledge used by economists is published. ``Cit.'' indicates the number of citations from citing journals to items in different types of outlets. ``Share'' indicates the fraction of citations to the items classified in each of the four outlet types relative to the total number of citations. The classification of the outlets is taken from  OpenAlex. The entry ``Books'' includes the OpenAlex types ``book series" and ``ebook platform''. The entry ``Repositories'' includes the OpenAlex types ``repository'' and ``other''.} 
\label{tab:outlet_types}
\end{table}

\subsection{Self-referentiality indicators observed at field level}

We start our analysis of self-referentiality at the most aggregate level. We consider economics as a unique set composed by all the articles published in the citing journals. Each reference is classified as journal self-reference, within-cluster, in-any-cluster, or within-economics. Four different shares are computed by calculating the proportion of these four classes of references relative to the total number of references. The four indicators of self-referentiality are synthesized in Table \ref{tab:self_ref_shares}.

% latex table generated in R 4.2.2 by xtable 1.8-4 package
% Fri Feb  6 17:24:43 2026
\begin{table}[ht]
\centering
\begin{tabular}{ccccc}
  \toprule
       & Journal         & Within & In any & Within  \\ 
Period & self-references & cluster & cluster & economics  \\ 
  \midrule
2006-2008 & 0.083 & 0.240 & 0.557 & 0.596   \\ 
  2012-2014 & 0.080 & 0.235 & 0.548 & 0.585   \\ 
  2019-2021 & 0.070 & 0.209 & 0.506 & 0.542   \\ 
   \bottomrule
\end{tabular}
\caption{Share of self-referentiality. Proportions, relative to the total number of references per period, of the following reference types:  (i) Journal self-references: references from a journal to articles in that same journal; (ii) Within-cluster: references from a journal of a cluster to other journals in the same cluster; (iii) In-any-cluster: references to journals classified in any of the clusters of the citing journals; (iv) Within-economics: references to EconLit economics journals.} 
\label{tab:self_ref_shares}
\end{table}

The first indicator is the share of journal self-references which diminished from 8.3\% of the pre-crisis period to 7\% of the last period. 

The \textit{within-cluster referentiality} diminished from 24\% of the pre-crisis period to 20.9\% of the last period, highlighting a general tendency toward a greater use of knowledge produced outside the own cluster.    

The \textit{in-any-cluster referentiality} indicator operates as if the set of clusters of the citing journals represent economics domain, thereby offering a first proxy of general disciplinary closure. The results indicate that just over half (approximately 50–55\%) of all references remain within this defined domain. Furthermore, the data confirm a contracting trend in this form of self-referentiality, with the share decreasing from 55.7\% in the first period to 50.6\% in the last.

The fourth indicator measures \textit{within-economics referentiality}. The share of references within economics started at approximately 60\% prior to the financial crisis, declining to 54.2\% in the most recent period. 

According to these four indicators, all the dimensions of self-referentiality in economics -- at the overall field level -- have shown a downward trend since the financial crisis. In particular, all indicators show a very small reduction between the first two periods and a more considerable reduction in the last period.

\subsection{Self-referentiality computed at cluster level}

The second step of the analysis consists in examining the data by aggregating journals according to their clusters. Each cluster is conceived as the set of articles from the journals belonging to that cluster. Once again, the journal as the unit of analysis is lost, and an aggregation of articles is performed.

This level of analysis enables two distinct types of investigation. The first focuses on the cited references within each cluster, which allows for an exploration of the knowledge base upon which the articles of the cluster are built. The second perspective examines the citations received by a cluster, tracing its impact both from within the cluster itself (intra-cluster citations) and from articles outside it (inter-cluster citations).

By comparing these two perspectives, we can evaluate several key aspects of a cluster. It allows us to assess its degree of self-referentiality, e.g. the extent to which it cites its own prior work. Furthermore, it helps to determine the weight of the knowledge produced within the cluster relative to the total knowledge it uses as seen in its cited references. Finally, this comparison provides insight into the use of the knowledge produced by a cluster by journals outside its boundaries, effectively measuring its external influence.

\begin{sidewaysfigure}
    \centering
    \begin{minipage}{0.9\textheight}  % 90% dell'altezza della pagina (che in landscape diventa la larghezza)
        \centering
        \begin{subfigure}[b]{0.45\linewidth}
            \includegraphics[width=\linewidth]{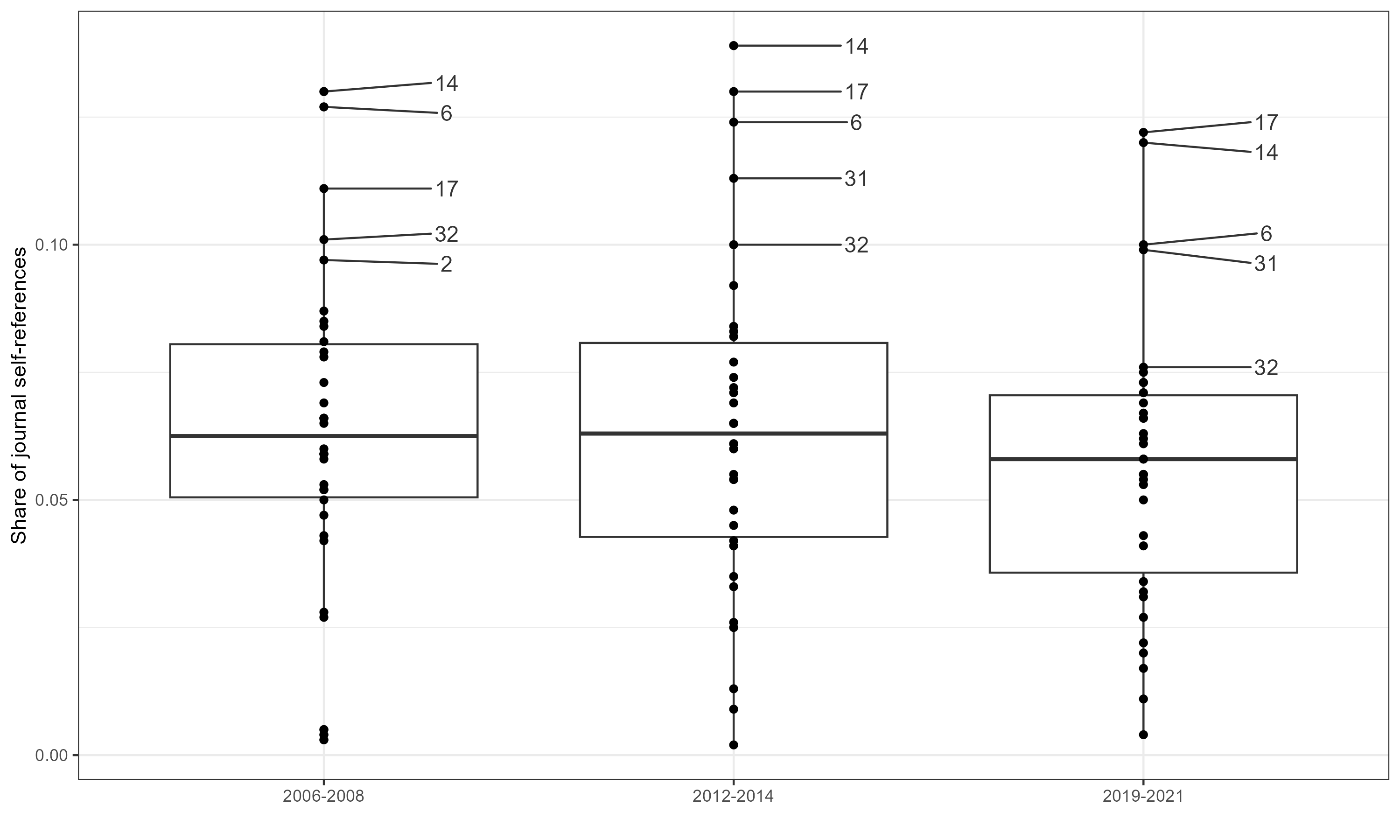}
            \caption{Share of journal self-references.}
            %\label{fig:primo}
        \end{subfigure}
        \hfill
        \begin{subfigure}[b]{0.45\linewidth}
            \includegraphics[width=\linewidth]{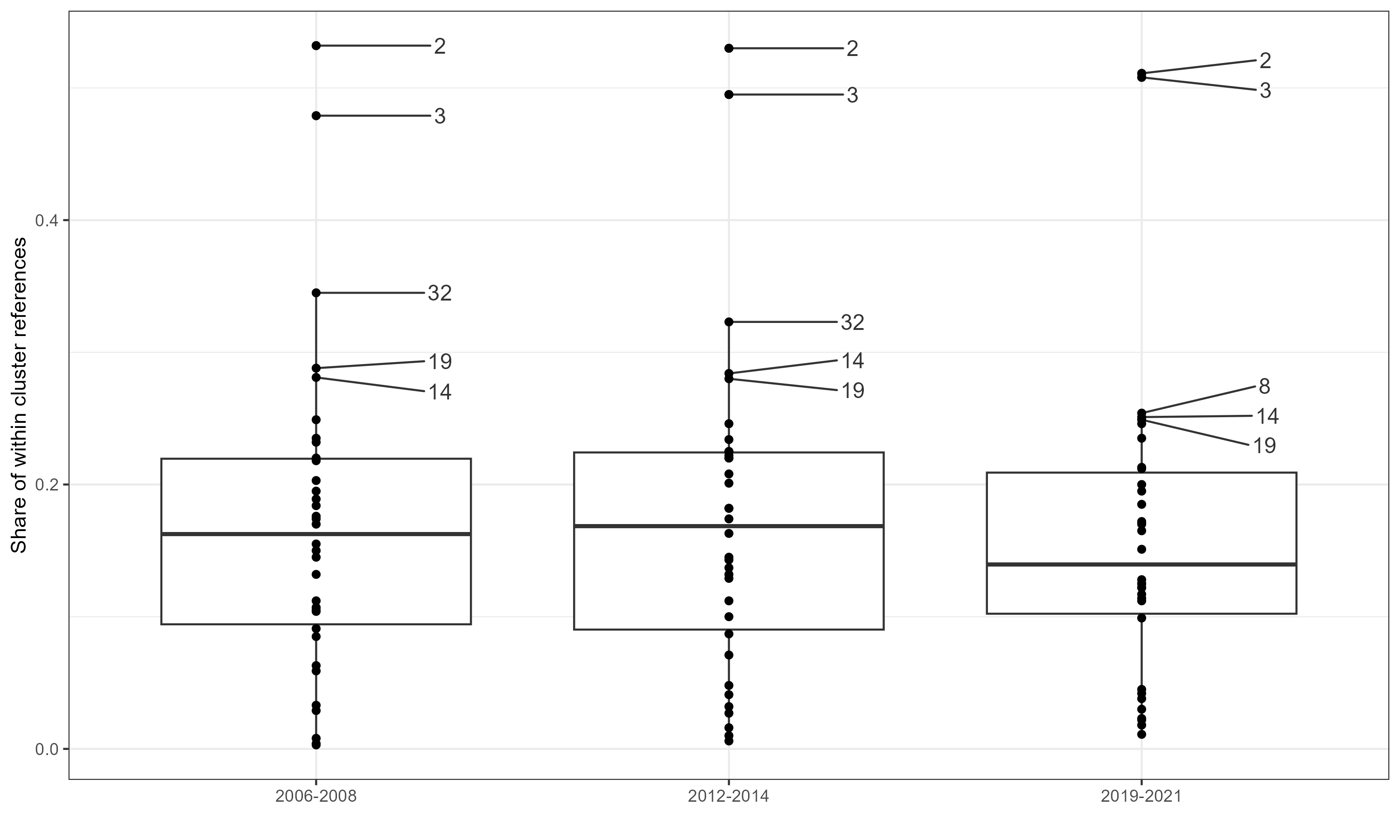}
            \caption{Share of within cluster references.}
            %\label{fig:secondo}
        \end{subfigure}
        
        \vspace{0.5cm}
        
        \begin{subfigure}[b]{0.45\linewidth}
            \includegraphics[width=\linewidth]{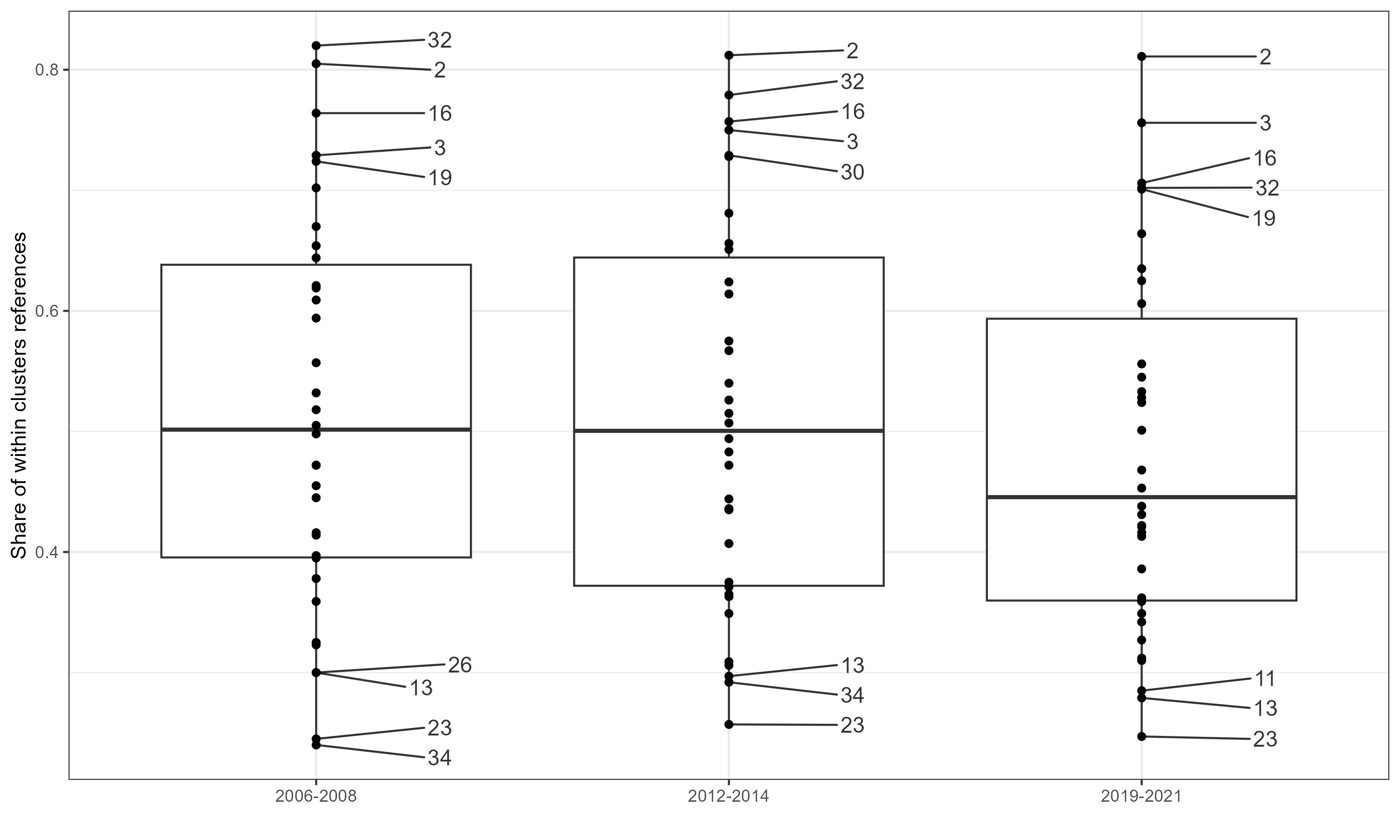}
            \caption{Share of in-any-cluster references.}
            %\label{fig:terzo}
        \end{subfigure}
        \hfill
        \begin{subfigure}[b]{0.45\linewidth}
            \includegraphics[width=\linewidth]{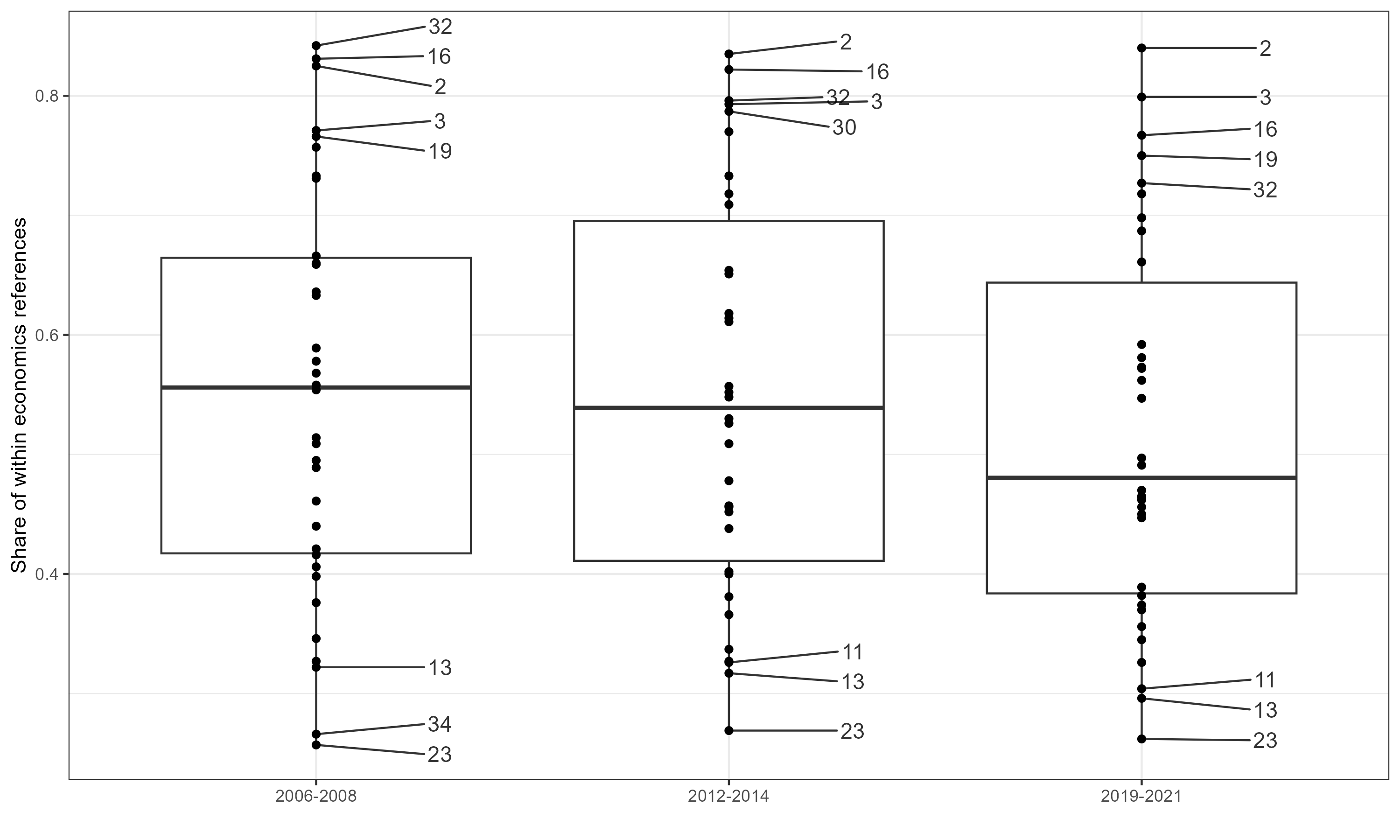}
            \caption{Share of within economics references.}
            %\label{fig:quarto}
        \end{subfigure}
        
        \caption{Self-referentiality of clusters of economics journals. For each cluster, proportions relative to the total number of cited references of the following reference types: (panel a) journal self-references; (panel b) within-cluster: references from journals of a cluster to other journals in the same cluster; (panel c) in-any-cluster: references to any of the clusters; (panel d) within-economics: references from a cluster to EconLit economics journals.}
        \label{fig:cluster_self_referentiality}
    \end{minipage}
\end{sidewaysfigure}

By adopting the first perspective and observing the cited references, we can reiterate the calculations of the four indicators of self-referentiality, this time for clusters. These indicators are reported in Figure \ref{fig:cluster_self_referentiality}. The figure clearly shows that the data calculated at the field level are the composition of very different behaviours across clusters. In particular, two clusters exhibit a very singular behaviour in terms of self-referentiality: they progressively tend to differentiate themselves from the other clusters. These clusters are Finance (cluster 2) and CORE (cluster 3), which at the end of the period are outliers in all self-referentiality dimensions except for journal self-references. While for the field as a whole, and for the most part of clusters, the share of self-referentiality decreases over time, these two clusters show an opposite trend, with a growing share of self-referentiality. The within cluster self-referentiality of these two cluster is more than double than the median value calculated for the other clusters. The shares of within clusters and within economics references is higher than $0.8$ against median values of $0.5$ or less, and it is growing, especially for the CORE, in the aftermath of the financial crisis.

\begin{figure}
    \centering
    \includegraphics[width=0.95\textwidth]{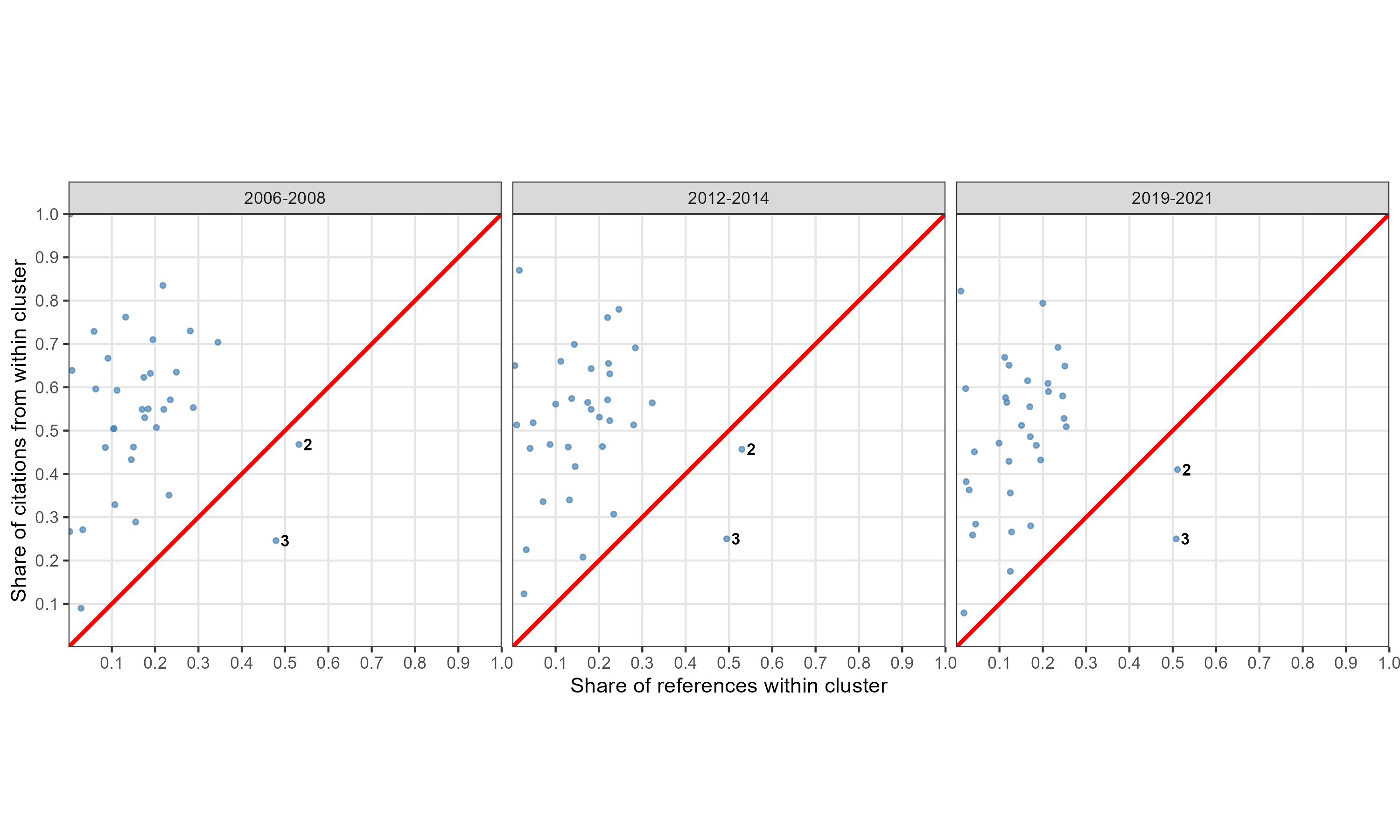}
    \caption{Within-cluster self-referentiality and external influence. The x-axis represents the share of references to journals within the same cluster, i.e. the degree of within-cluster self-referentiality; the y-axis represents the share of citations that each cluster receives from within the cluster: the lower the share, the higher the external influence generated by the knowledge produced in the cluster.}
    \label{fig:self_ref_vs_influence}
\end{figure}

Figure \ref{fig:self_ref_vs_influence} illustrates the relationship between within-cluster self-referentiality and its external influence. The $x$-axis measures the within cluster self-referentiality, and the $y$-axis measures the share of citations that each cluster receives from within the cluster: a lower share indicates that the knowledge produced in the cluster is used more extensively outside of it, thus signaling greater influence on other economics clusters. The three distribution of Figure \ref{fig:self_ref_vs_influence} do not change over time. In fact, a multivariate 3-sample statistical test for the equality of bivariate distributions was computed in the \proglang{R} computing environment \citep{RN13} using the \code{eqdist.etest} function in the package \pkg{energy} \citep{RN29}. It does not reject the null hypothesis that the bivariate distributions of the clusters are the same across the three periods ($E$-statistic $=$ $0.374$, $p = 0.97$, based on $9,999$ permutations).

From a descriptive point of view, the two outlier clusters exhibit distinctive patterns over the three periods: the cluster 2 (Finance) moves slighlty  downwards and to the left, while the cluster 3 (CORE) shifts slightly to the right. Regarding the share of self-citations from within the cluster, Cluster 2 (Finance) shows values similar to many other clusters and a slight tendency toward reduction, from 0.468 in the first period to 0.410 in the last. This indicates a tendency toward a growing external influence over time. Cluster 3 (CORE), on the other hand, has the second-lowest value in the first period and shows a very slight tendency toward growth, from 0.246 to 0.250, that is a very slight tendency toward the reduction of its influence on other economics clusters. As for the share of references within the cluster, the two clusters have very similar values, and they are the highest. Finance shows a tendency toward the reduction of within-cluster self-referentiality over time, decreasing from 0.532 to 0.511, while CORE shows a tendency toward growth, increasing from 0.479 to 0.508.

\subsection{Reference asymmetry at cluster level}

The analysis has thus far examined self-referentiality in its various declinations, revealing significant variation across clusters, with Finance and the CORE emerging as outliers. We now extend the analysis to explore inter-cluster relationships, investigating the extent to which each cluster draws upon knowledge produced in other clusters. 

\begin{sidewaysfigure}
    \centering
    \includegraphics[width=0.95\textwidth]{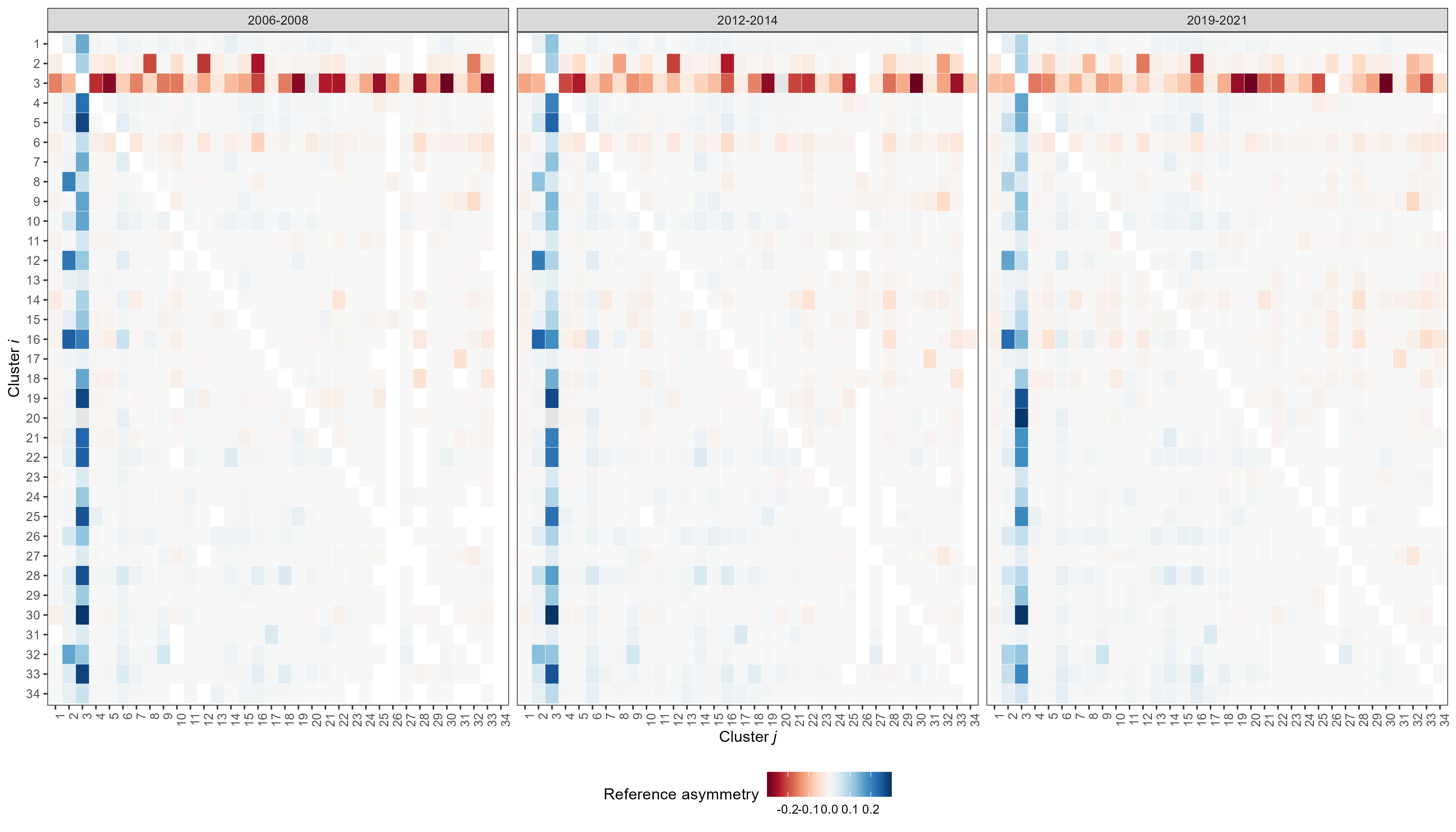}
    \caption{Heatmaps of the Reference Asymmetry (RA) index between clusters. Due to the antisymmetric property of the RA index, heatmap cells $(i, j)$ and $(j, i)$ exhibit opposite signs and opposite colors, while sharing the same color intensity, which reflects the magnitude of the asymmetry. In the heatmap, reading across rows, a positive RAI indicates that the row cluster relies more intensively on knowledge from the column cluster than vice versa; a negative value indicates the opposite. Reading down columns, the interpretation is reversed: a positive value means that the column cluster relies more intensively on knowledge from the row cluster, and a negative value means the opposite.}
    \label{fig:RA_index}
\end{sidewaysfigure}

The RA index is illustrated in Figure \ref{fig:RA_index}: the heatmap displays the RA values for all pairs from our 34 clusters across the three periods. Clearly, in each period, the majority of pairs exhibit values around symmetry, indicating similar incoming and outgoing flows.  
A striking pattern emerges for the CORE (3): all values in its row are negative. This indicates that the CORE consistently relies less intensively on knowledge from every other cluster than those clusters rely on knowledge from it. In other words, the CORE functions as a net source of knowledge across all bilateral relationships. Examining the temporal evolution of the CORE's bilateral asymmetries, most clusters exhibit a tendency toward greater mutual balance with the CORE, with negative values progressively weakening. Two clusters Microeconomics (19) and Public economics, Public Finance (30) confirm over time their pronounced negative asymmetries, indicating that their reliance on the CORE is stable. Labour Economics (20) presents the most dramatic shift: initially the only cluster with balanced exchanges with the CORE in the first period, it ends with a strongly negative asymmetry, reflecting a marked increase in its dependence on knowledge from the CORE.

The Finance cluster (2) exhibits a more nuanced pattern in its row. Its RA values are negative for most other clusters, though with a milder intensity compared to the CORE, indicating a moderate tendency to rely less on their knowledge than they rely on its own. The highest intensities are with Accounting cluster (8), and the other two finance-related clusters (12 and 16). The only exception is the pair with the CORE, where the value is positive, meaning that Finance draws more intensively on knowledge from the CORE than vice versa. Over time, these negative asymmetries show a gradual reduction in magnitude, suggesting a trend toward greater mutual balance in Finance's bilateral relationships.

The row for Econometrics and Statistics (6) displays very mild negative asymmetries with almost all other clusters. This pattern reflects the instrumental and methodological nature of its scientific output, which is drawn upon relatively evenly across clusters. The only exception is again the CORE, where the relationship shows a positive asymmetry, indicating that Econometrics and Statistics relies more intensively on knowledge from the CORE than vice versa. 

Consider the other general cluster (4), which collects Post-Keynesian and heterodox journals. Its row shows a positive, though slightly decreasing, asymmetry with the CORE. This can be conjectured as originating from the necessity to critique the main tenets of the CORE, and confirms the asymmetry found in specific research areas such as DSGE models and macro agent-based models \citet{crossfertil}. The only cluster with which Post-Keynesian economics exhibits a negative asymmetry is HET and Methodology (cluster 25), reflecting the interest of historians of economic thought and methodologists in the heterodox traditions.

\subsection{Self-referentiality indicators computed at journal level}

\begin{sidewaysfigure}
    \centering
    \begin{minipage}{0.8\textheight}  % 90% dell'altezza della pagina (che in landscape diventa la larghezza)
        \centering
        \begin{subfigure}[b]{0.45\linewidth}
            \includegraphics[width=\linewidth]{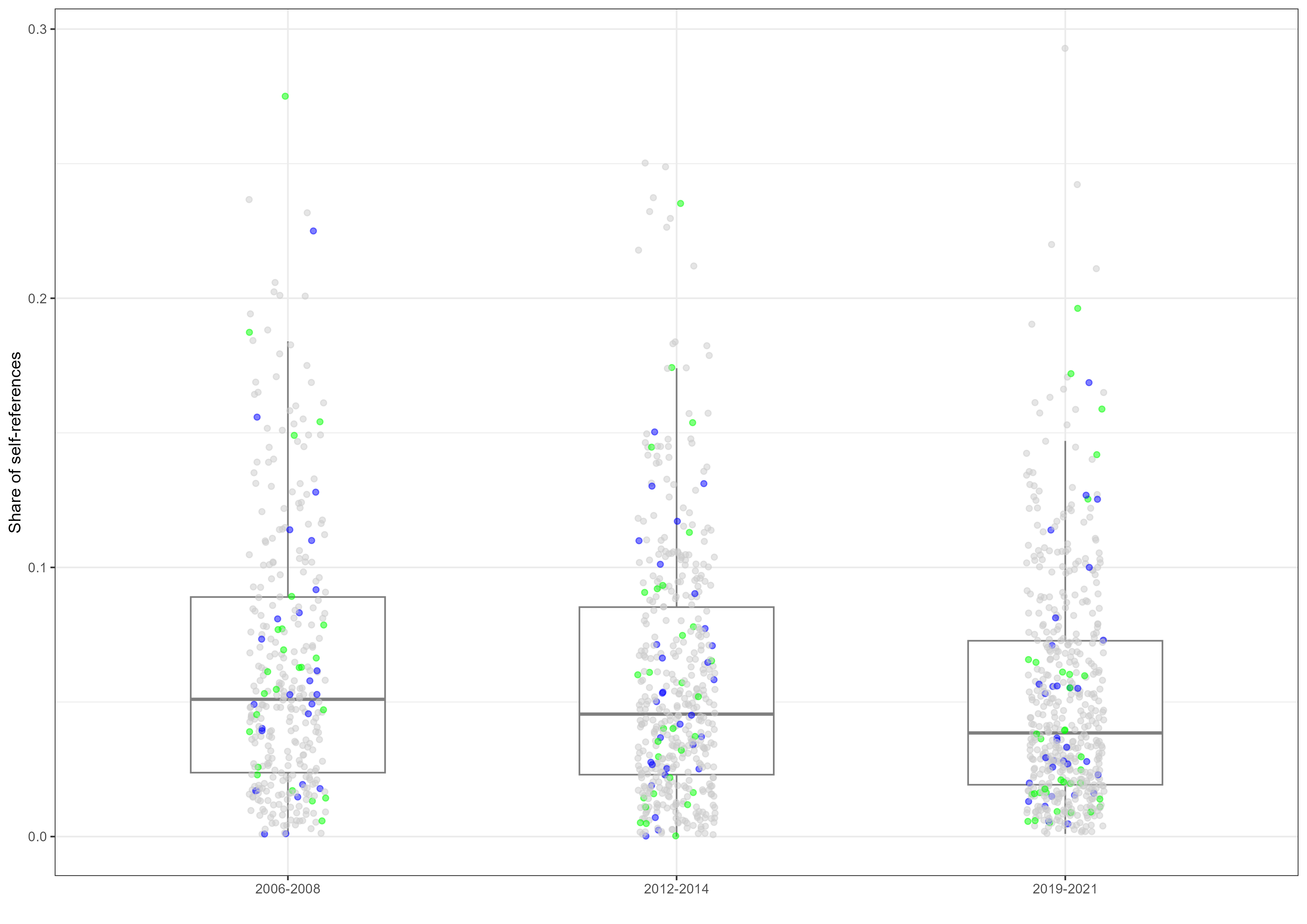}
            \caption{Journal self-references.}
            \label{fig:primo}
        \end{subfigure}
        \hfill
        \begin{subfigure}[b]{0.45\linewidth}
            \includegraphics[width=\linewidth]{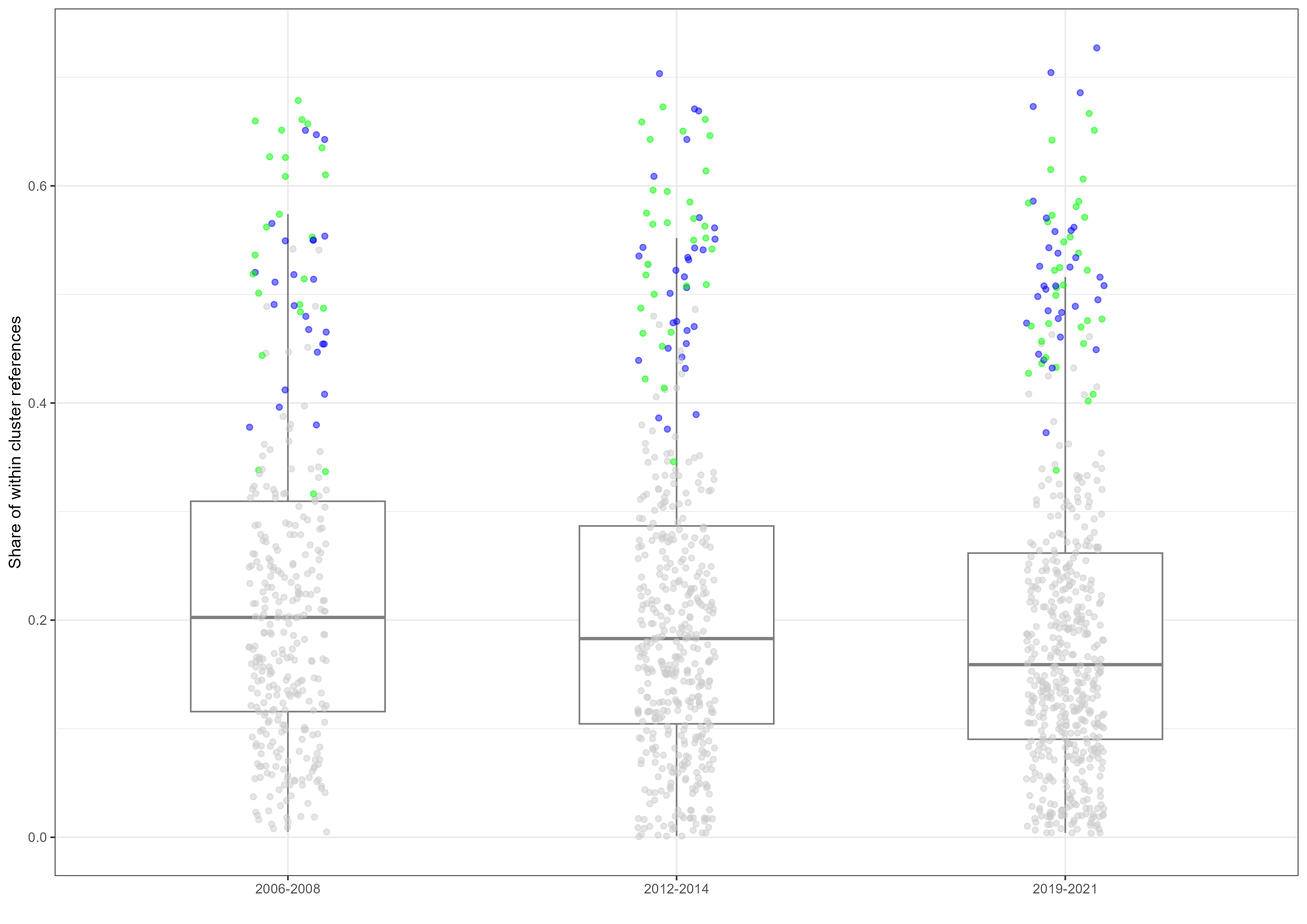}
            \caption{Share of within cluster references.}
            \label{fig:secondo}
        \end{subfigure}
        
        \vspace{0.5cm}
        
        \begin{subfigure}[b]{0.45\linewidth}
            \includegraphics[width=\linewidth]{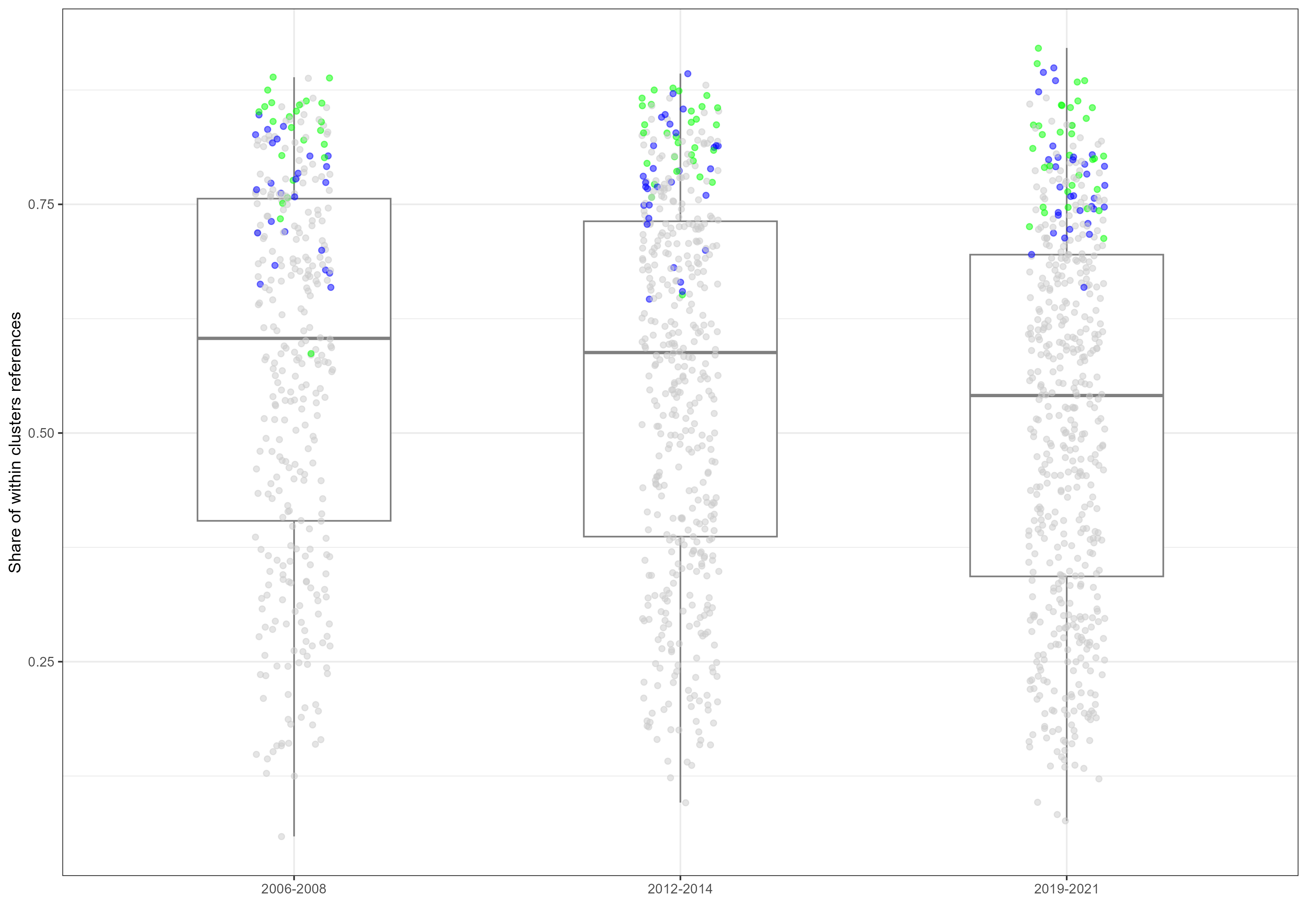}
            \caption{Share of in-any-cluster references.}
            \label{fig:terzo}
        \end{subfigure}
        \hfill
        \begin{subfigure}[b]{0.45\linewidth}
            \includegraphics[width=\linewidth]{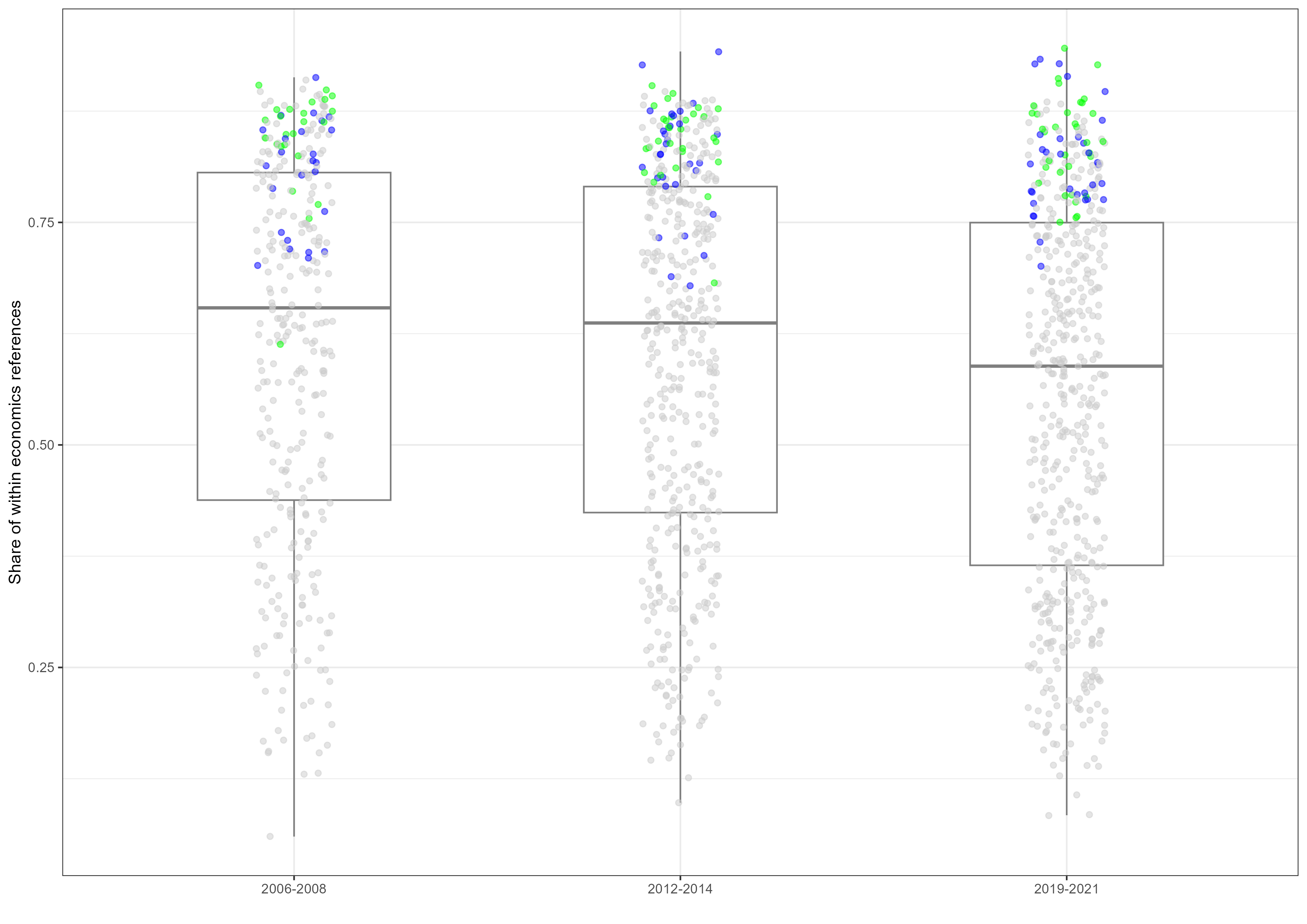}
            \caption{Share of within economics references.}
            \label{fig:quarto}
        \end{subfigure}
        
        \caption{Journal self-referentiality. For each journal, proportions relative to the total number of cited references of the following reference types: (panel a) journal self-references; (panel b) within-cluster: references from a journal to other journals in the same cluster; (panel c) in-any-cluster: references from a journal to journals classified in the considered clusters; (panel d) within-economics: references from a journal to EconLit economics journals. Green points represent journals of Cluster 2 (Finance); blue points are journals of Cluster 3 (CORE).}
        \label{fig:journal_self_referentiality}
    \end{minipage}
\end{sidewaysfigure}

The next step of analysis consists in considering the self-referentiality indicators at journal level. Figure \ref{fig:journal_self_referentiality} reports the distributions of journals according to the four indicators of self-referentiality. It details the composition of bibliographic references for each journal, showing the proportions of: journal self-references (panel a); references to other journals within the same cluster (panel b); references to journals classified in the citing clusters (panel c); and references to economics journals indexed in EconLit (d). As discussed earlier regarding the cluster-level analysis, two outlier groups emerge, and these are highlighted in color: the green points represent the journals of the Cluster 2 (Finance), while the blue points correspond to those of Cluster 3 (CORE). This color coding allows for a direct comparison of citation patterns between these two distinctive groups.

Figure \ref{fig:journal_self_referentiality} highlights that the estimates of the median values of the indicators are sensitive to the level of disaggregation adopted. For instance, when the median share of within-clusters references is calculated as the median of journal shares, it tends to be higher than when the data are calculated by computing the median of shares calculated for clusters.

A second point is relevant. When journal self-references are observed, the journals of the two outlier clusters, that is the clusters of Finance and CORE, appear to have shares that are very dispersed over the entire distribution. When, instead, the other indicators of self-referentiality are observed, the journals of the two clusters appear all near or outside the upper bound of the inter-quartile range, with the highest values of the whole set of journals. Moreover, all journals appear to have a trend toward a growing self-referentiality according to all three dimensions considered.

\begin{sidewaysfigure}
    \centering
    \includegraphics[width=0.95\textwidth]{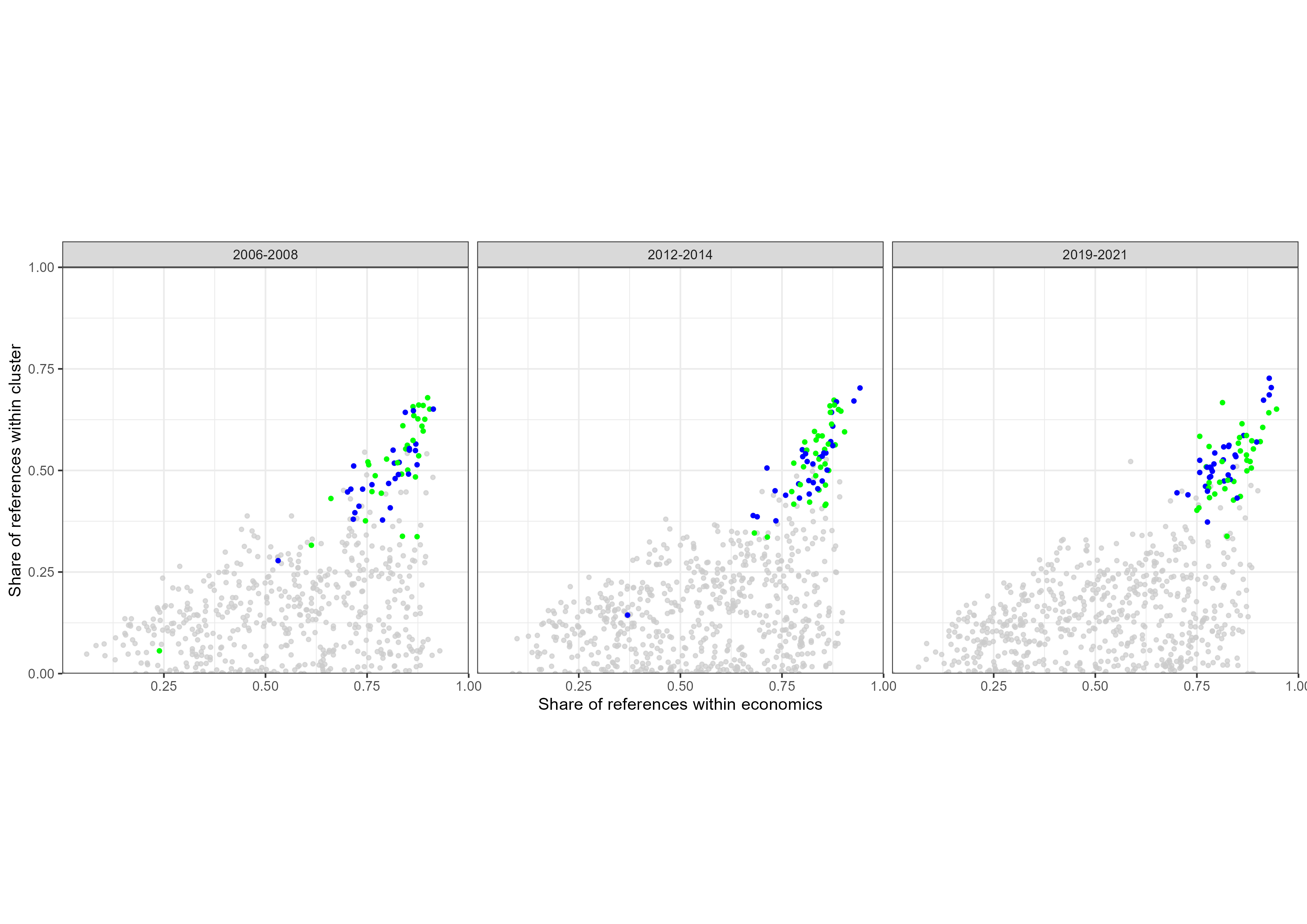}
    \caption{Distribution of journals according to the use of self-referential knowledge. The x-axis represents the share of references to economics journals; the y-axis represents the share of references from a journal in a cluster to other journals within the same cluster. Green points represent journals of Cluster 2 (Finance); blue points are journals of Cluster 3 (CORE).}
    \label{fig:scatter_knowledge_base}
\end{sidewaysfigure}

A better understanding of the use of the knowledge base can be achieved by combining, for each journal, the share of within-cluster and within-economics references. Figure \ref{fig:scatter_knowledge_base} shows, again, that the journals of Finance and CORE are more self-referential than all the other journals in economics.  Their references are drawn much more from within cluster and from within economics compared to all the journals in the other clusters. The visual inspection also clearly indicates that the journals of both clusters exhibit growing self-referentiality, both within their cluster and within economics, over time. A multivariate 3-sample statistical test for the equality of bivariate distributions was computed in the \proglang{R} computing environment \citep{RN13} using the \code{eqdist.etest} function in the package \pkg{energy} \citep{RN29}. It rejects the null hypothesis that the bivariate distributions of the journals are the same across the three periods ($E$-statistic $=$ $6.94$, $p < 0.001$, based on $9,999$ permutations).

\begin{figure}
    \centering
    \includegraphics[width=0.95\textwidth]{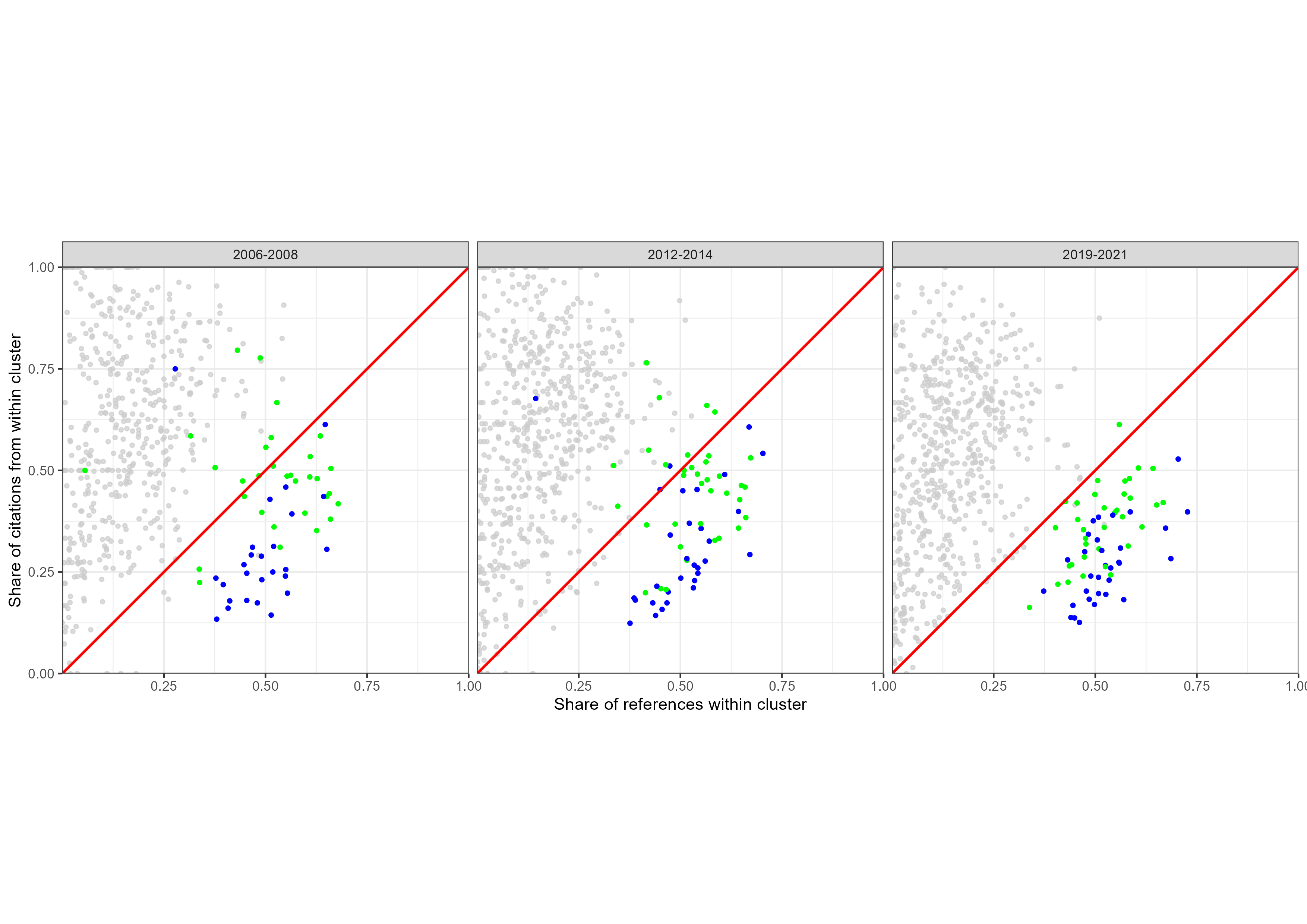}
    \caption{Journal within-cluster self-referentiality and external influence. The x-axis represents the share of references from a journal to other journals within the same cluster, i.e. journals degree of within-cluster self-referentiality; the y-axis represents the share of citations that each journal  receives from journals of its same cluster: the lower the share, the higher the external influence of a journal over other economic journals. Green points represent journals of Cluster 2 (Finance); blue points are journals of Cluster 3 (CORE).}
    \label{fig:scatter_cit_vs_know_clus}
\end{figure}

In order to explore the relationship between self-referentiality and external impact at journal level, Figure \ref{fig:scatter_cit_vs_know_clus} develops at journal level what we have seen for clusters in Figure \ref{fig:self_ref_vs_influence}. The x-axis measures the share of references that each journal makes to journals within the same cluster. The y-axis indirectly measures the external influence of journals as the share of citations that each journal receives from other journals of its same cluster. In this case, lower shares indicate that the knowledge produced by a journal is used more extensively by journals outside its cluster, thus signaling greater external influence. Figure \ref{fig:scatter_cit_vs_know_clus} confirms the distinctive patterns of Cluster 2 (Finance) and Cluster 3 (CORE) when they are observed at the journal level. In Finance, the group of journals that initially had a low level of external influence —located to the left of the bisector— have tended to reduce the share of citations received from within the cluster, converging toward the journals that were positioned to the right of the bisector from the start. Meanwhile, no clear modification in the share of references is visible. This is coherent with the data calculated at the cluster level where it was visible a slight growth of external influence, i.e. reduction of the share of citations from within the cluster, and a very small reduction of within cluster self-referentiality. The journals of the Finance cluster show decreasing over time shares of citations received from within the cluster, by indicating a growing influence in economics and a high level of self-referentiality. The journals of the CORE cluster display, for the most part, low and decreasing over time shares of citations received from inside the cluster. These two clusters, especially the CORE, appear to have stable and relatively high influence over other clusters in economics, combined with the highest and growing level of within cluster self-referentiality.

\newpage

\subsection {Reference asymmetry at journal level}

The reference asymmetry index computed at the journal level helps complete the picture. When reference asymmetries of journals belonging to the same cluster are computed, internal asymmetries emerge in many cases. These reveal that a few journals, often only one or two, export knowledge to all the other journals in the cluster, i.e., they exhibit negative row-wise asymmetries. This is the case, for instance, of the \textit{Journal of Economic Theory} in the Microeconomic Theory cluster (19), of \textit{World Development} in the Development cluster (18), of the \textit{Journal of Human Resources} in Labour Economics (20), and of the \textit{RAND Journal of Economics} in Industrial Organization (21). 

\begin{sidewaysfigure}
    \centering
    \includegraphics[width=0.95\textwidth]{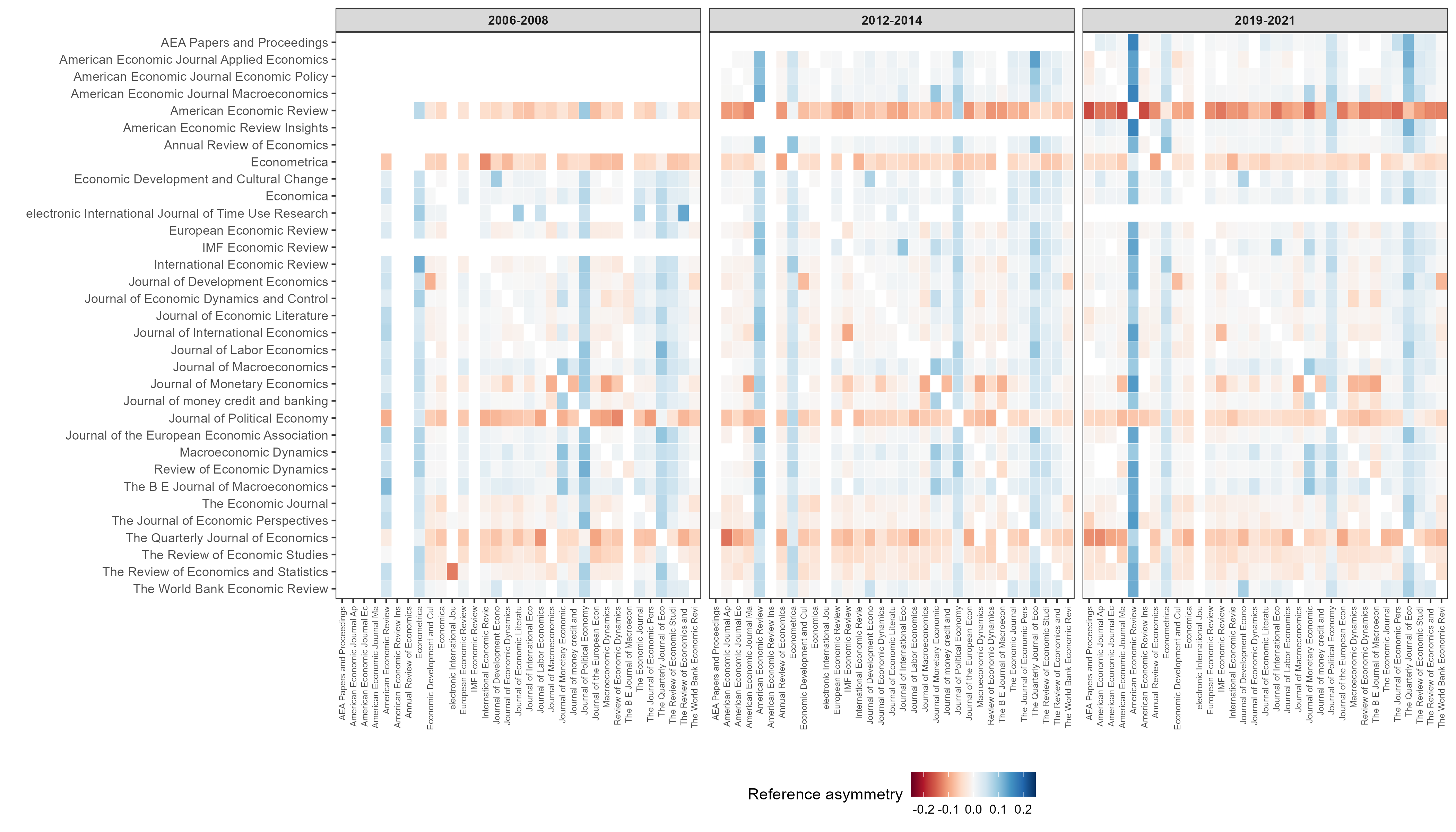}
    \caption{Heatmaps of the Reference Asymmetry (RA) index between journals of the CORE. Due to the antisymmetric property of the RA index, heatmap cells $(i, j)$ and $(j, i)$ exhibit opposite signs and opposite colors, while sharing the same color intensity, which reflects the magnitude of the asymmetry. Reading across rows, a negative RA indicates that the row journal relies less intensively on knowledge from the column journal than vice versa; a positive value indicates the opposite. Reading down columns, the interpretation is reversed. A Shiny app is available Reference Asymmetry index for the whole dataset at: https://goeld.shinyapps.io/j\_asym\_app/}
    \label{fig:RAI_CORE}
\end{sidewaysfigure}

Finance and the CORE confirm as exceptions also in this respect. In Finance, the \textit{Journal of Finance} shows the strongest reference asymmetries with respect to all the other journals, but three other journals are characterized by a prevalence of reference asymmetries, namely the \textit{Review of Financial Studies}, the \textit{Journal of Financial Economics}, and the \textit{Journal of Banking and Finance}. The CORE, as reported in Figure \ref{fig:RAI_CORE}, is characterized by negative asymmetries linking the leading journals, i.e. the top-5 journals of economics, with all the other journals in the cluster. The top-5 export knowledge to the other journals; in particular, they export knowledge to second-tier generalist journals and to the field-specific top journals within the CORE. 

When the reference asymmetry is computed between the journals of the CORE and the journals of the other clusters, a pattern emerges clearly. The relationships are systematically asymmetric, with the journals of the CORE exporting knowledge to the journals of the other clusters. Furthermore, the Top-5 are always characterized by negative asymmetry with respect to all the other clusters; however, negative asymmetries also emerge for specific top-field journals within the CORE when compared to the journals of their respective field clusters. The \textit{Journal of Development Economics} and \textit{Economic Development and Cultural Change} exhibit negative asymmetries with respect to almost all other journals in the Development Economics cluster (18); the \textit{Journal of Labor Economics} shows negative asymmetry with all journals in the Labor Economics cluster. And, as can be expected, \textit{Econometrica} displays strong negative asymmetry with almost all journals in the Econometrics and Statistics cluster. There are very few exceptions to this pattern; the most relevant are the three leading journals of the Finance cluster mentioned above and the \textit{Journal of Economic Theory} in the Microeconomic Theory cluster.

\section {Discussion}

The analysis of self-referentiality yields different results depending on the level of aggregation considered. At the aggregate level, economics journals as a whole show a decline in self-referentiality across all dimensions, pointing to a general trend toward greater openness. According to these four indicators, all dimensions of self-referentiality in economics have exhibited a downward trend since the financial crisis. In particular, all indicators show a very small reduction between the first two periods and a more considerable reduction in the most recent one. 

However, when a finer level of granularity is introduced, the picture becomes more complex. We adopted scientific journals as the main unit of observation. They are considered as institutionalized spaces where social networks and cognitive content intersect. It is within these spaces that the discipline becomes more visible and traceable over time, and that the questions of self-referentiality and reference asymmetry can be framed. 

We have examined journals at two interconnected levels. At the meso level, we analyze aggregations of journals into clusters based on the classification developed in our previous work \citep{baccini2025}. This classification is developed by fusing information on intellectual similarity (references used and content) and social proximity (authors who publish in the journals and editors who govern them). At the individual level, each journal is treated both as a standalone unit and as a member of a specific cluster, allowing us to move between micro and meso levels of analysis.

The analysis at the cluster level identified two clusters, namely Finance and CORE, that display divergent behavior relative to the others throughout the entire period. As already remarked, the CORE cluster concentrates both general-interest journals and the top journals of specialized fields whose remaining outlets are located in other specialized clusters. Analogously the Finance cluster collects top-tier journal of Finance \citep{baccini2025}. Both clusters exhibit high levels of self-referentiality across all dimensions, emerging as clear outliers over the whole period, with the sole exception of journal self-reference, which remains in line with other clusters. While the Finance cluster shows a consistent reduction in self-referentiality over time, the CORE exhibits the opposite trend, moving toward increasing closure.

When the clusters are observed from the point of view of Reference Asymmetry, that is by considering the flows of knowledge between them, Finance and the CORE differentiate again from all the other clusters, but showing different characteristics from each other. 

Finance shows strongly negative asymmetric relationships with the other finance-related clusters and weakly negative asymmetric relationships with a majority of the other clusters. These relationships appear stable over time. This indicates that knowledge produced in Finance is generally used by other clusters, especially by other finance-related clusters, more than finance uses knowledge produced in those areas. However, the relationship with the CORE cluster is inverted: Finance draws on knowledge produced in the CORE more than CORE draw on knowledge produced in Finance. In sum, Finance operates as a net knowledge exporter, except in its relationship with the CORE, where it becomes a net importer. 

The CORE, instead, is characterized by strong negative reference asymmetry with all the other clusters. The asymmetry is particularly pronounced with respect to clusters representing the main traditional body of economics and the most policy relevant ones: i.e. Microeconomics, Labour economics, and Public finance/Public economics. Over time, these asymmetric relationships show a tendency to strengthen with these three clusters, while exhibiting a weak tendency to diminish with all the others. This pattern reveal that that the CORE acts as a net knowledge exporter toward the other clusters, especially toward those that have traditionally formed the bulk of economics. More strikingly, these results permit to transform the map of proximity between journals developed by \citet{baccini2025}, into a picture of knowledge dependence flows. Over the entire period, the CORE occupies a position of hierarchical centrality in knowledge flows, shaping what constitutes as valid knowledge within economics. 

When data are analyzed at the maximum level of granularity represented by individual journals, the results observed at the cluster level are reinforced. All journals belonging to the two outlier clusters are themselves outliers with respect to the other journals. All these journals are characterized by high levels of self-referentiality, with the sole exception of journal self-reference. A clear divergence emerges over time between CORE and Finance journals: Finance journals show a generalized tendency toward reduction of self-referentiality across all dimensions, whereas CORE journals move in the opposite direction. \citet{Truc_2023} showed that Top-5 economics journals exhibit a trend opposite to that of other economics journals, namely a reduction in their degree of interdisciplinarity, that is the complement of self-referentiality. The present analysis adds that all journals within the CORE cluster, not only the Top-5, move in the same direction, toward increasingly higher levels of self-referentiality.

As for reference asymmetry, hierarchically ordered knowledge flows emerge both within and between clusters. Within almost all clusters, including Finance and CORE, one or more journals display reference asymmetries with other journals in the same cluster, meaning that a small set of journals exports knowledge to the rest of the cluster. As for flows between journals of different clusters, the analysis shows that journals in the CORE and Finance clusters export knowledge toward journals in the other clusters. Reference asymmetries between journals of different clusters are rare, and the transmission of knowledge appears to be mediated primarily by the CORE. In 2006-2009, at the apex of this hierarchically ordered flow of knowledge were the \textit{Journal of Finance} and the \textit{Journal of Financial Economics} of the Finance cluster, and \textit{Econometrica}, the \textit{Journal of Political Economy}, the \textit{Economic Journal},the \textit{Quarterly Journal of Economics} and the \textit{American Economic Review} of the CORE. Notably, the \textit{Journal of Finance} and the \textit{Journal of Financial Economics} imported knowledge from the other journals at the hierarchical apex of economics. In 2019-2021, the journals of the apex are the same, but the \textit{American Economic Review} assumes a leadership position in exporting knowledge.

Up to this point, the focus had been exclusively on the intellectual dimension, examining the self-referentiality and asymmetry of references at the level of clusters and clustered journals. However, the clusters we used for the analysis were constructed taking into account the interconnections between the academic communities of publishers and authors that gravitate around them. In the specific case of the CORE and Finance clusters, the main driver of their internal cohesion is interlocking editorship, i.e., the cross-presence of the same editors in the boards of different journals \citep{baccini2025}. Editors act as gatekeepers of these journals \citep{baccini_re} and, by virtue of the hierarchically ordered knowledge flows, effectively control the main flows of knowledge within discipline. As we have seen, Finance and the CORE differ precisely in their knowledge export patterns: while the CORE exports knowledge to all other clusters, including Finance, Finance exports knowledge only toward other second-tier finance clusters. In this sense, and in accordance with \citet[p. 105]{Fourcade}, finance appears as an autonomous discipline with its own rules and scholarly community, yet one that remains intellectually indebted to the CORE of economics. This asymmetry reinforces the hierarchical structure of economics. 

\section{Conclusions}

Our results suggest that economics is organized around a central group of journals — the CORE — whose gatekeepers define what counts as legitimate knowledge and who is entitled to produce it. In turn, within the CORE a internal hierarchy exists: generalist journals dominate field-specific top journals, which in turn depend on generalists more than vice-versa. Hence, the CORE gatekeepers have leverage over which legitimate knowledge can be channeled toward other clusters of journals, who in turn are substantially unable to reciprocate. This one-way dependency is not merely intellectual in nature; it also shapes the recruitment and career progression systems in economics, which are based primarily on access to the most prestigious journals \citep{heckman2020}. This multidimensional dependence is possibly stronger within the CORE cluster itself, and in the clusters forming the traditional bulk of economics (microeconomics, labour economics and public economics/public finance). It likely weakens with peripheral clusters, where intellectual dependence is less structured and career dynamics are probably influenced also by rules defined within the peripheral clusters themselves. 

The analysis of changes intervened after the 2008 financial crisis, has highlighted that the high self-referentiality of the CORE has been increasing. This has been accompanied by a strengthening of reference asymmetry toward the clusters that represent the historical bulk of economics. This result appears consistent with an interpretation of the post-2008 financial crisis response as a form of epistemic closure of economics: a defense of the discipline's methods and fields against external criticism \citep{mirowski, aignerkapeller}. This pattern of intellectual closure finds its organizational counterpart in the long-term dynamics of editorial boards documented by \citet{baccini2026}. After the crisis, the organizational structure of the top economics journals has turned inward,
consolidating incumbent journals and their gatekeepers while curtailing the entry to top positions of new journals and scholars. As the CORE tightened its intellectual grip on the traditional fields of economics, it simultaneously consolidated its gatekeeping structures, limiting organizational diversity.
If instead we look at economics at large, including more peripheral subfields, we find evidence of the opening process documented by other authors. The \textit{kind} of exchange this openness implies, however, is yet to be explored and is likely a mixture of genuine intellectual interaction and mere export of applied econometric techniques compatible with the phenomenon of economic imperialism.

In sum, our analysis reveals a publishing system in which intellectual and organizational closure at the top of the hierarchy have reinforced one another. Peripheral areas of the discipline, on the other hand, have higher rates of interdisciplinary interaction, which possibly signals a decline in the ability of top journals to act as a faithful representation of the overall discipline.

\section*{Declarations}

\begin{itemize}
\item Funding: the research is funded by the Italian Ministry of University, PRIN project: How economics is changing: A multilayer network analysis of the recent evolution of economics journals, between specialization and self-referentiality (1980-2020),  2022SNTEFP, PI: Alberto Baccini.
\item Conflict of interest/Competing interests: The authors have no competing interests to declare that are relevant to the content of this article.
\item Data, source code, including the code of the Shiny app, and supplementary materials necessary to reproduce the analysis are permanently archived on Zenodo at https://doi.org/10.5281/zenodo.19403867. The interactive Shiny app is available at https://goeld.shinyapps.io/j\_asym\_app/. 
%\item Preprint: the article is available at %\url{https://arxiv.org/pdf/2305.00026.pdf}
\item Authors' contributions: Alberto Baccini and Carlo Debernardi contributed to the study conception and design, data collection, data analysis and visualization. Both participated to the writing of the manuscript.
\end{itemize}

\bibliography{references}  %%% Uncomment this line and comment out the ``thebibliography'' section below to use the external .bib file (using bibtex) .

\




% \end{thebibliography}

\newpage

\section*{Appendix A}
\renewcommand{\thesection}{A.\arabic{section}}
\setcounter{section}{0}
\renewcommand{\thetable}{\Alph{section}.\arabic{table}}
\setcounter{table}{0}

\section{Robustness check: self-referentiality indicators under alternative definitions of economics journals}

In this type of analysis, as noted by \citet{Truc_2023}, varying definitions of the set of economics journals can yield different results. To assess the robustness of our findings, we employ two external definitions of economics journals: a more restrictive one based on \citet{Truc_2023}'s journal list, and a broader one derived from OpenAlex classification. The comparison of the sizes of the three journal sets are reported in Table \ref{tab:n_journals}. We then replicate the aforementioned indicators using these alternative definitions.

% latex table generated in R 4.2.2 by xtable 1.8-4 package
% Mon Feb  9 15:02:34 2026
\begin{table}[ht]
\centering
\begin{tabular}{lrrrrrr}
  \toprule
   & \multicolumn{2}{c}{EconLit} & \multicolumn{2}{c}{Truc et al.} & \multicolumn{2}{c}{OpenAlex} \\
   \cmidrule(lr){2-3} \cmidrule(lr){4-5} \cmidrule(lr){6-7}
       & Citing & Cited & Citing & Cited & Citing & Cited \\ 
  \midrule
2006-2008 & 571 & 799 & 176 & 318 & 287 & 1034 \\ 
  2012-2014 & 686 & 988 & 190 & 342 & 329 & 1378 \\ 
  2019-2021 & 665 & 1064 & 189 & 345 & 320 & 1578 \\ 
   \bottomrule
\end{tabular}
\caption{Economics journal counts under three classification schemes. Comparison between: (1) our EconLit-based dataset; (2) \citet{Truc_2023}'s journal list; (3) OpenAlex journals categorized as ``Economics, Econometrics and Finance'. Counts of citing (cited) journals derive from our initial citing (cited) journal set, classified according to the other two schemes.}
\label{tab:n_journals}
\end{table}

The list of economics journals used in \citet{Truc_2023} is more restrictive than EconLit list. Table \ref{tab:truc} reports the self-referentility indicators computed by adopting Truc et al.'s classification. The share of self-references calculated for this list of journals is substantially identical to that calculated for EconLit journals. The other indicators tend to have slightly lower values than those calculated for EconLit journals, although the downward trend from the first to the last period is confirmed. These latter values can be considered for each period as the lower bound limits of self-referentiality within the discipline. 

% latex table generated in R 4.2.2 by xtable 1.8-4 package
% Tue Feb 17 17:34:10 2026
\begin{table}[ht]
\centering
\begin{tabular}{ccccc}
  \toprule
            & Journal & Within & In any & Within \\ 
    Period & self references & cluster & cluster & economics  \\ 
  \midrule
2006-2008  & 0.082 & 0.242 & 0.543 & 0.591  \\ 
  2012-2014  & 0.083 & 0.232 & 0.518 & 0.560  \\ 
  2019-2021  & 0.071 & 0.205 & 0.466 & 0.502  \\ 
   \bottomrule
\end{tabular}
\caption{\citet{Truc_2023}'s list of economics journals: share of self-referentiality in . Proportions, relative to the total number of references per period, of the following reference types: (i) Truc et al.'s journal self-references: references from a Truc et al. journal to articles in that same journal; (ii) Within-cluster: references from a Truc et al.'s journal of a cluster to other Truc et al.'s journals in the same cluster; (iii) In any cluster: references to Truc et al.'s journals classified in any of the clusters of the Truc et al.'s citing journals; (iv) Within-economics: references to Truc et al.'s economics journals.} 
\label{tab:truc}
\end{table}

For OpenAlex classification, we consider as economics journals those having  a predominant focus on economics. Across the three periods examined, less than half of EconLit citing journals have a predominant focus on ``Economics, Econometrics and Finance'' in OpenAlex. Conversely, the set of cited journals with a predominant focus on ``Economics, Econometrics and Finance'' in OpenAlex is larger than the list of EconLit cited journal.

% latex table generated in R 4.2.2 by xtable 1.8-4 package
% Mon Feb  9 12:13:17 2026
\begin{table}[ht]
\centering
\begin{tabular}{ccccc}
  \toprule
            & Journal & Within & In any & Within \\ 
    Period & self references & cluster & cluster & economics  \\ 
  \midrule
2006-2008  & 0.071 & 0.265 & 0.615 & 0.740  \\ 
  2012-2014 & 0.064 & 0.251 & 0.600 &  0.741 \\ 
  2019-2021 & 0.054 & 0.217 & 0.552  &  0.710 \\ 
   \bottomrule
\end{tabular}
\caption{OpenAlex definition of economics journals: share of self-referentiality. Proportions, relative to the total number of references per
period, of the following reference types: (i) OA journal self-references: references from a OpenAlex journal to articles in that same journal; (ii) Within-cluster: references from a OpenAlex economics journal journal of a cluster to other OpenAlex journals in the same cluster; (iii) In any cluster: references to OpenAlex journals classified in any of the
clusters of the OpenAlex citing journals; (iv) Within-economics: references to OpenAlex economics journals.} 
\label{tab:OpenAlex}
\end{table}

Table \ref{tab:OpenAlex} reports the indicators computed by adopting OpenAlex classification.
\begin{comment}
The column ``EconLit citing OA'' indicates the share of references from Econlit journals to economics journals classified according to OpenAlex. For the three periods the shares of references to OpenAlex economics journals are lower than the ones calculated adopting EconLit classification for cited journals, despite the list of cited economics journals is bigger in the first case. This is probably due to the fact than OpenAlex classification consider as ``Economics, Econometrics and Finance'' many journals receiving few citations and exclude journals receiving many citations from EconLit journals, such as business and statistics journals.    
\end{comment} 
OpenAlex journals show a slightly lower share of self-references than EconLit journals, though the trends are consistent. In contrast, all other indicators of self-referentiality are higher than those calculated for EconLit journals, even though the downward trend over time is respected. In particular, self-referentiality within economics is $10$ to $15$ percentage points higher when using the definition of economics journals provided by OpenAlex. These latter values can be considered for each period as the upper bound limits of self-referentiality within the discipline.

Results computed at the field level indicate that the self-referentiality indicators based on EconLit journals fall somewhere between those calculated using stricter and broader definitions of economics journals. All the data suggest a tendency toward a reduction in the discipline's self-referentiality in the aftermath of the financial crisis. More precisely, this reduction becomes more pronounced in the most recent period.

\end{document}